\documentclass[final,3p,times,twocolumn]{elsarticle}

\usepackage{amssymb}
 \usepackage{amsthm}
 \usepackage{amsmath}
 {
      
      \newtheorem{theorem}{\bf{Theorem}}[section]
      \newtheorem{definition}{\bf{Definition}}

      \newtheorem{remark}{\bf{Remark}}

       \newtheorem{proposition}{\bf{Proposition}}
       
}
\DeclareMathOperator*{\argmin}{arg\,min}
\usepackage[monochrome]{xcolor}

\usepackage{graphics}
\graphicspath{ {./image/} }
\usepackage{algorithm}

\journal{Control Engineering Practice}

\begin{document}

\begin{frontmatter}



\title{Moving-horizon False Data Injection Attack Design against Cyber-Physical Systems}

\author[label1]{Yu Zheng \footnote{yzheng6@fsu.edu}}
\author[label1]{\textcolor{blue}{Sridhar Babu Mudhangulla}}
\author[label1]{Olugbenga Moses Anubi}
 \affiliation[label1]{organization={Department of Electrical and Computer Engineering},
             addressline={Florida State University},
             city={Tallahassee},
             postcode={32303},
             state={FL},
             country={USA}}

\begin{abstract}
Systematic attack design is essential to understanding the vulnerabilities of cyber-physical systems (CPSs), to better design for resiliency. In particular, false data injection attacks (FDIAs) are well-known and have been shown to be capable of bypassing bad data detection (BDD) while causing targeted biases in resulting state estimates. However, their effectiveness against moving horizon estimators (MHE) is not well understood. In fact, this paper shows that conventional FDIAs are generally ineffective against MHE. One of the main reasons is that the moving window renders the static FDIA recursively infeasible. This paper proposes a new attack methodology, moving-horizon FDIA (MH-FDIA), by considering both the performance of historical attacks and the current system's status. Theoretical guarantees for successful attack generation and recursive feasibility are given. Numerical simulations on the IEEE-14 bus system further validate the theoretical claims and show that the proposed MH-FDIA outperforms state-of-the-art counterparts in both stealthiness and effectiveness. In addition, \textcolor{blue}{an experiment on} a path-tracking control system of an autonomous vehicle shows the feasibility of the MH-FDIA in real-world nonlinear systems. 

\end{abstract}




\end{frontmatter}


\section{Introduction}
The control of CPSs is generally a closed-loop feedback process, in which the physical processes, measured by distributed sensors, are driven by control actions depending on the accurate estimates of the state variables. However, the cyber layers are vulnerable to adversarial attacks, and the closed-loop interaction propagates the effect of the inevitable attacks on the physical processes. Recent examples, such as the malicious attacks on Israel's water supply system \cite{brentan2021cyber}, the 2015 Ukraine blackout \cite{7752958} and the recent leakage of the Colonial Pipeline due to cyberattacks \cite{reeder2021cybersecurity}, \textcolor{blue}{indicate} that cyberattacks can cause severe consequences for CPS stakeholders. Developing resilient CPSs in an adversarial setting has motivated researchers to study possible attack strategies, such as denial of service (DoS) \cite{xu2006jamming} and deception attacks \cite{pasqualetti2013attack}.

False data injection attack (FDIA), most widely studied deception attack, follows a general attack strategy of maximizing the effectiveness on the system behaviors while maintaining stealthiness \cite{pasqualetti2013attack}. Stealthiness measures the potential to bypass the bad data detection (BDD) \cite{hu2018state}, and effectiveness measures the closeness to the intended degradation of system performance \cite{zhang2020false}. Early researchers incorporated the full system model into the maximization program to generate a feasible attack \cite{liu2011false, mo2010false, hu2018state}. This rendered the resulting process computationally inefficient and less pragmatic FDIAs. The authors in \cite{liu2011false} studied the FDIA generation problem against the least-square estimator (LSE) with a residual-based BDD. The feasibility of FDIA against the Kalman filter with $\chi^2$ detector was studied in \cite{mo2010false}. A sufficient and necessary condition for insecure estimation under FDIA was derived for the networked control system in \cite{hu2018state}. To develop more pragmatic attack generation strategies, several constraints are incorporated to capture the attacker limitations such as limited access to sensors \cite{liu2011false}, incomplete knowledge of system dynamics \cite{liu2015modeling}, and incomplete knowledge of implemented state estimators \cite{lu2022false}. Data-driven approaches have also been used to generate FDIA in order to incorporate constrained knowledge of system dynamics. Examples include supervised learning using existing small attack dataset \cite{mohammadpourfard2020generation}, physics-guided unsupervised learning \cite{Zheng2021Algorithm, khazraei2022learning}. Some recent results have streamlined their approach to generate feasible attack that maintains stealthiness and ignored the effectiveness maximization objectives, largely due to the inevitable nonconvexity which results in computationally ineffective FDIAs at best \cite{sui2020vulnerability}.


In this paper, we consider an additional perspective for the FDIA generation problem: How does the attack history affect the feasibility of the FDIA problem at the current time? In other words, is recursive feasibility essential for FDIA \textcolor{blue}{problem}? Moving-horizon estimation (MHE) \cite{rawlings2017model} has been widely used in different control systems and is being deployed in CPSs with linear/nonlinear dynamics, networked and distributed architectures \cite{zou2020moving}. One of the main design focuses of MHE is how to guarantee feasibility over the next window given the solution in the current window. This is called recursive feasibility \cite{lofberg2012oops, muske1993receding}. From this angle, MHE is inherently more resilient than static counterparts since any successful FDIA against MHE must themselves be recursively feasible. In this paper, we refer to estimators designed with a window size of $1$ as ``static state estimators (SSE)", and the FDIA designed to target those estimators as ``static FDIA". One of the main differences between SSE and MHE is in their consideration of recursive feasibility. We show that static FDIAs have a low chance of success against BDD built on MHE. Consequently, we propose a more general framework for moving-horizon FDIA (MH-FDIA) design with consideration of recursive feasibility.

\textbf{Contribution}: As a result of the ineffectiveness of static FDIAs against MHE, this paper studies a novel and more challenging, but more pragmatic FDIA problem against MHE. \textcolor{blue}{The paper discusses the limitations of various representative FDIA designs against MHE and proposes an adaptive mechanism for generating provably successful MH-FDIAs with consideration of recursive feasibility.} The proposed solution solves the problem directly without any heuristic simplification or transformation. As a result, the proposed framework is general for any combination of MHE and BDD considerations. Theoretical results \textcolor{blue}{are} validated based on an IEEE 14-bus system and a nonlinear path-tracking control system of a wheeled mobile robot. \textcolor{blue}{Notably}, to the best knowledge of the authors, this paper is the first study of MH-FDIA with consideration of recursive feasibility.

The remainder of the paper is organized as follows. In Section~\ref{Sec:notation}, the notations employed are summarized. In Section~\ref{Sec:model}, a concurrent system model of the CPS, including linear physical model, MHE, and BDD, is given for the subsequent development of the proposed FDIA scheme. In Section~\ref{Sec:MHFDIA}, an MH-FDIA generation scheme is given as a feasibility problem, and an adaptive algorithm is used to produce all feasible FDIAs. In Section~\ref{Sec:Simulation}, the state-of-the-art MH-FDIA generation algorithm is compared to a directed applied eigenvalue maximization approach. The comparison is done via numerical simulation on an IEEE 14-bus system \textcolor{blue}{and an experiment on a wheeled mobile robot}. Conclusions follow in Section~\ref{Sec:Conlusion}.

\section{Preliminary}\label{Sec:notation}
We use $\mathbb{R}, {\mathbb R}^n, {\mathbb R}^{m \times n}$ to denote Euclidean spaces of real scalars, $n$-dimensional  \textcolor{blue}{column} vectors, and $m \times n$ matrix respectively. \textcolor{blue}{Normal-face lower-case letters $(e.g.\hspace{1mm} x \in {\mathbb R})$ are used to represent real scalars, bold-face lower-case letters $(e.g.\hspace{1mm} \mathbf{x} \in {\mathbb R}^n)$ represent vectors, while normal-face upper-case letters $(e.g.\hspace{1mm} X \in {\mathbb R}^{m \times n})$ represent matrices.} $X^{\top}$ and $X^{\dagger}$ denote the transpose and left pseudo-inverse of matrix $X$ respectively \textcolor{blue}{$(i.e. X^{\dagger}X=I, (I-XX^{\dagger} )X = 0)$.} Let $X \in \mathbb{R}^{m \times n}$ $(m<n)$ be a full-ranked matrix, $X^{\perp}$ denotes the orthogonal complement of $X$ (i.e. $X^{\top}X^{\perp} =0$ and $[X \hspace{2mm} X^{\perp}]$ is full-ranked). Let $\mathcal{T} \subseteq \{1,\dots, m\}$ be a set of indices, then for a matrix $X \in {\mathbb R}^{m \times n}$, $X_{\mathcal{T}}\in {\mathbb R}^{|\mathcal{T}| \times n}$ is a sub-matrix with the rows of $X$ corresponding to the indices in $\mathcal{T}$. We use $\rho(A)$ to denote the spectral radius of the matrix $A$. 

\textcolor{blue}{We use $\mathbf{x}_i$ and $\mathbf{x}(i)$ to denote the vector $\mathbf{x}$ at time instance $i$ and the $i$-th elements of vector $\mathbf{x}$ respectively.} $I_T \triangleq [i-T+1,i-T+2,\cdots, i]$ denotes a moving-window of fixed size $T$ \textcolor{blue}{at time instance $i$}. The subscript $T$ will be omitted if clear from the context. Accordingly, $I_T+1 (\text{or} \hspace{1mm} I+1) \triangleq [i-T+2,i-T+3, \cdots, i+1]$ denotes the $T$ time window from $i-T+2$ to $i+1$, and \textcolor{blue}{$I^{-} \triangleq [i-T+1,i-T+2,\cdots, i-1]$} denotes a history window of size $T-1$.
$\mathbf{x}_I = [\mathbf{x}_{i-T+1}^{\top}, \mathbf{x}_{i-T+2}^{\top}, \cdots, \mathbf{x}_{i}^{\top}]^{\top} \in \mathbb{R}^{Tn}$ is a \textcolor{blue}{column} vector composed of vector \textcolor{blue}{$\mathbf{x}_j \in \mathbb{R}^n$} for all $j \in I$. If a continuous function $\phi:[0,a] \rightarrow [0,\infty)$ is monotonously increasing and satisfies $\phi(0)=0$, then we say $\phi$ belongs to class $\kappa$. The complement of a set $\mathcal{S}$ is denoted by $\bar{\mathcal{S}}$ (or ${\mathcal{S}}^c$). The support of a vector \textcolor{blue}{$\mathbf{x} \in \mathbb{R}^n$ is defined as
$ \textsf{supp}(\mathbf{x}) \triangleq \{i\subseteq \{1,\dots, n\}|\mathbf{x}(i) \neq 0\}
$.} $\Sigma_{k} \triangleq \{\mathbf{x} \in \mathbb{R}^n \lvert |\textsf{supp}(\mathbf{x})| \leq k\}$ denotes the set of $k$-sparse vectors. $\mathbf{x}_I \in \Sigma_k$ means all component vectors are $k$-sparse. Given a matrix $X \in \mathbb{R}^{m \times n}$ and a support $\mathcal{T}$, we say $X_{\mathcal{T}} \in \Sigma_k$ if each row of $X_{\mathcal{T}}$ belongs to $\Sigma_k$. We use the symbol $\otimes$ to denote the Kronecker product, i.e. for $B \in \mathbb{R}^{m\times n}$,
$$ \begin{bmatrix}
    a_{11} & a_{12}\\
    a_{21} & a_{22}
\end{bmatrix} \otimes B = 
\begin{bmatrix}
     a_{11}B & a_{12}B\\
    a_{21} B & a_{22}B
\end{bmatrix} \in \mathbb{R}^{2m \times 2n}.
$$.

\section{Problem formulation}\label{Sec:model}
\textcolor{blue}{Consider a nonlinear model used to describe the underlying physical processes of CPS
\begin{equation}\label{equ:nonlinear_sys}
\begin{aligned}
    \dot{\mathbf{x}} &= f(\mathbf{x},\mathbf{u})\\
     \mathbf{y}&=g( \mathbf{x}) + \mathbf{v}
     \end{aligned}
\end{equation}
where $\mathbf{x} \in \mathbb{R}^n, \mathbf{y} \in \mathbb{R}^m, \mathbf{v} \in \mathbb{R}^m$ are the internal state variables, sensor measurements, and noise. $f$ and $g$ are assumed lipschitz in $\mathbf{x}$ and $\mathbf{u}$. Given an equilibrium point $(\mathbf{x}_0, \mathbf{u}_0)$ and a fixed time step\footnote{\textcolor{blue}{The physical processes interact with the cyber components operating in discrete time intervals, so researchers often use discrete model for studies of control and estimation in CPS \cite{weerakkody2019resilient}.}} $T_s$, a discrete linear time-invariant (LTI) model is used to approximate the dynamics of the physical plant around the equilibrium point:
\begin{equation}
    \begin{aligned}
        \Delta{\mathbf{x}}_{i+1} &= A^{\prime}\Delta{\mathbf{x}}_i + B^{\prime}\Delta{\mathbf{u}}_i,\\
    \mathbf{y}_i &= C\Delta{\mathbf{x}}_i+\mathbf{v}_i  ,
    \end{aligned}
\end{equation}
where $\Delta{\mathbf{x}}_{i} = \mathbf{x}_i-\mathbf{x}_0$, $\Delta{\mathbf{u}}_{i} = \mathbf{u}_i-\mathbf{u}_0$, and
$$A^{\prime} = \frac{\partial f}{\partial \mathbf{x}}\bigg|_{(\mathbf{x}_0, \mathbf{u}_0)} T_s + I_n, \hspace{2mm} B^{\prime} = \frac{\partial f}{\partial \mathbf{u}}\bigg|_{(\mathbf{x}_0, \mathbf{u}_0)} T_s, \hspace{2mm}C = \frac{\partial g}{\partial \mathbf{x}}\bigg|_{\mathbf{x}_0}. $$
Consider a stable controller $\Delta{\mathbf{u}}_i = K\Delta{\mathbf{x}}_i$, where $K$ is designed properly to achieve $\underset{i \rightarrow \infty}{\lim} \|\Delta{\mathbf{x}}_i\| = 0$. Then we study the attack design for the system at the stable equilibrium points. Consequently, the following closed-form dynamical model is used throughout the subsequent development:
}
\begin{equation}\label{equ:sys_model}
\begin{aligned}
    \mathbf{x}_{i+1} &= A\mathbf{x}_i,\\
    \mathbf{y}_i &= C\mathbf{x}_i+\mathbf{v}_i  ,  
\end{aligned}
\end{equation}
\textcolor{blue}{where $A = A^{\prime} + B^{\prime}K$ is stable. For the sake of clean presentation, we make the abuse of $\mathbf{x}_i$ as $\Delta \mathbf{x}_i$ in the model \eqref{equ:sys_model}.} Then the measurement model on the window $I$ is given by
\begin{equation}\label{equ:meas_model_windowI}
    \mathbf{y}_I = H\mathbf{x}_i+\mathbf{v}_I,
\end{equation}
where $H=\begin{bmatrix}CA^{1-T}\\CA^{2-T}\\ \vdots \\C \end{bmatrix}$ is a backward observation matrix and \textcolor{blue}{$\mathbf{y}_I = [\mathbf{y}_{i-T+1}^{\top} \hspace{2mm} \mathbf{y}_{i-T+2}^{\top} \hspace{1mm}\cdots \hspace{1mm}\mathbf{y}_{i}^{\top}]^{\top} \in \mathbb{R}^{mT}$}, $\mathbf{v}_I \in \mathbb{R}^{Tm}$ are the horizontal measurement vector and the noise vector\textcolor{blue}{, respectively,} in the window $I$. The following widely used assumptions are made for the system model above:
\begin{enumerate}
    \item The CPS \eqref{equ:sys_model} is asymptotically stable: $0<\rho(A)<1$.
    \item The pair $(C,A^{-1})$ is observable: $\textsf{rank}(H)=n$.
    \item The noise is bounded: $\|\mathbf{v}_I\|_2\leq \varepsilon_v$, for a known constant $\varepsilon_v > 0$.
\end{enumerate}

\noindent Consequently, \textcolor{blue}{we start to discuss the MHE and bad data detector (BDD) used in this paper. Given the measurement model in \eqref{equ:meas_model_windowI},} MHE seeks to obtain an estimate of the state vector $\mathbf{x}_i$ from the most recent $T$ measurements in the window $I$. 
\begin{definition}
[Moving-horizon estimator, MHE \cite{rao2003constrained,allan2019moving}]\label{def:MHE}
A moving-horizon estimator is an operator of the form $\mathcal{D}:\mathbb{R}^{Tm} \rightarrow \mathbb{R}^n$ which returns an estimate of the state vector from $T$-horizon observation. Moreover, $\mathcal{D}$ is said to be stable if there exists $\tau_0, \varepsilon_0 <\infty$ such that
\begin{equation}\label{equ:decoder}
    \|\mathcal{D}(\mathbf{y}_I) - \mathbf{x}_i\|_2 \leq \tau_0 \|\mathbf{x}_i\|_2 + \varepsilon_0,
\end{equation}
for all $i>0$. 

\end{definition}
\begin{remark}
 For $\ell_2$ MHE, $\mathcal{D}_2$ is a linear function of $\mathbf{y}_I$ and given by:
\begin{equation}\label{equ:L2_decoder}
\mathcal{D}_2(\mathbf{y}_I) \triangleq \argmin_{\mathbf{x}}\|\mathbf{y}_I-H\mathbf{x}\|_2 = H^{\dagger}\mathbf{y}_I. 
\end{equation}
\end{remark}

\noindent \textcolor{blue}{As MHE estimates the states from the measurements of window $I$, BDD monitors the state estimation process to detect any malicious inputs.} It is often designed based on the residual $\|\mathbf{y}_I - H\mathcal{D}(\mathbf{y}_I)\|_p, \hspace{1mm} 0<p\leq \infty$. $p=2$ is commonly used \cite{liu2011false, mo2010false}. $p=1$ and $p=\infty$ have also been used \cite{kosut2010limiting}. The BDD considered in this paper is based on the \textcolor{blue}{$\ell_2$ MHE in \eqref{equ:L2_decoder}.}
\begin{definition}[Bad data detector, BDD]
Consider the MHE in \eqref{equ:L2_decoder} and the measurement model in \eqref{equ:meas_model_windowI}. Given a threshold $\delta>0$, a bad data detector is a discriminator given by 
\begin{align}\label{equ:BDD}
    \textsf{BDD}(\mathbf{y}_I) = \left\{\begin{array}{cl}1&\text{ if }\left\|\mathbf{y}_I-H\mathcal{D}_2(\mathbf{y}_I)\right\|_2 > \delta,\\0&\text{ otherwise.}\end{array}\right.
\end{align}
\end{definition}
Next, we discuss the ineffectiveness of static FDIA designs \cite{pasqualetti2013attack, liu2011false} against the MHE in \eqref{equ:L2_decoder} and the basic BDD in \eqref{equ:BDD}.
\begin{enumerate}
    \item The original FDIA design strategy is to define the \textcolor{blue}{attack $\mathbf{e}_i$} on the range space of the measurement matrix $C$, i.e. $\mathbf{e}_i = C\mathbf{a}_i$, where $\mathbf{a}_i$ is an arbitrary signal. We call this type of FDIA a perfect stealth attack against static $\ell_2$ estimator since the detection residual is zero:
    $$ \|\mathbf{y}_i + \mathbf{e}_i - CC^{\dagger}(\mathbf{y}_i + \mathbf{e}_i)\|_2 = \|(I-CC^{\dagger})C(\mathbf{x}_i+\mathbf{a}_i)\|_2 = 0.
    $$
    \textcolor{blue}{Here we ignore the noise $\mathbf{v}_i$ in the model \eqref{equ:sys_model}.} However, under the MHE, $\mathbf{y}_I = H\mathbf{x}_i+(I_T \otimes C)\mathbf{a}_I$, where \textcolor{blue}{$\mathbf{a}_I = \begin{bmatrix}\mathbf{a}_{i-T+1}^{\top}, & \cdots, & \mathbf{a}_i^{\top}\end{bmatrix}^{\top}$,} then the residual is 
    \begin{align}
        \left\|\mathbf{y}_I-H\mathcal{D}_2(\mathbf{y}_I)\right\|_2 &= \|(I-HH^{\dagger})\mathbf{y}_I\|_2  \nonumber \\
    &= \|(I-HH^{\dagger})(I_T \otimes C)\mathbf{a}_I\|_2 \nonumber
    \end{align} 
    \textcolor{blue}{Notice the second equality follows from $(I-HH^{\dagger})H = 0$.} It is clear that $\textsf{span}(I_T \otimes C) \cap \textsf{range}(H) \neq \emptyset$ is not guaranteed. In fact, if $H$ has full column rank, the above holds if and only if $A=\beta I$ for some $\beta \in \mathbb{R}$. In other words, static FDIA is effective against MHE if and only if the states of CPS are completely decoupled.
    
    \item Another well-known design is a generalized stealth FDIA \cite{liu2011false} which is given by the following program:
    \textcolor{blue}{
    \begin{equation}\label{equ:generalized_FDIA}
    \begin{aligned}
     \underset{\mathbf{e}_i}{\textsf{Maximize:}}& \hspace{2mm} \|\mathcal{D}_2(\mathbf{y}_i)-\mathcal{D}_2(\mathbf{y}_i + \mathbf{e}_i)\|_2, \\
    \textsf{Subject to:} & \hspace{2mm} \|\mathbf{y}_i + \mathbf{e}_i - C\mathcal{D}_2(\mathbf{y}_i + \mathbf{e}_i) \|_2 \leq \delta/T.
    \end{aligned} 
\end{equation}}
Considering the system model in \eqref{equ:sys_model} \textcolor{blue}{without noise}, the constraint in \eqref{equ:generalized_FDIA} is equivalent to $ \|(I-CC^{\dagger})\mathbf{e}_i\|_2 \leq \delta/T$ which does not guarantee that $\|(I-HH^{\dagger})\mathbf{e}_I\|_2 \leq \delta$.

\item Consider the case where the static design is done using $H$ instead of $C$, for example, $e_I = H\mathbf{a}_I$. Here the conflict arises when the time window moves to $I+1$. The design has to guarantee that the attack vector is consistent from window to window. This renders the static design infeasible for the moving-horizon estimator.
\end{enumerate}

Consequently, a moving-horizon FDIA design is needed against MHE with consideration for recursive feasibility.

\section{Moving-horizon FDIA design}\label{Sec:MHFDIA}
In this section, we start with a definition of successful FDIA, then develop the proposed MH-FDIA. The recursive feasibility of the moving-horizon schemes is also analyzed. The following assumptions on the attacks are widely used \cite{liu2011false,mo2010false}:
\begin{enumerate}
    \item The attacker has limited access to sensors; $\mathbf{e}_i \in \Sigma_k$ \textcolor{blue}{for some $k < m$}.
    \item The attacker can inject arbitrary values on compromised sensors.
    \item The attacker will not abandon compromised sensors or gain access to more sensors at runtime. Thus, the attack support is time-invariant ($\dot{\mathcal{T}}=0$).
\end{enumerate}
Henceforth, we denote the support of the attack sequence for the time-horizon $I$ by $\mathcal{T}_I = [\mathcal{T}^{\top}, m+\mathcal{T}^{\top}, \cdots, (T-1)m+\mathcal{T}^{\top}]^{\top}$, and $\mathcal{T}_{I^-}$ is defined accordingly. Next, we formalize the attack performance criteria used in this paper. 
\begin{definition}[successful FDIA \cite{sui2020vulnerability, zheng2021resilient}]
Given the estimator-detector pair $\left(\mathcal{D}_2(\mathbf{y}_I),\textsf{BDD}(\mathbf{y}_I)\right)$ in \eqref{equ:L2_decoder} and \eqref{equ:BDD} respectively, the attack vector $\mathbf{e}_I \in \Sigma_k$ is said to be $(\alpha,\epsilon)$-successful if the following hold:
\begin{equation}\label{equ:succ_FDIA}
    \|\mathcal{D}_2(\mathbf{y}_I)-\mathcal{D}_2(\mathbf{y}_I+\mathbf{e}_I)\|_2 \geq \alpha, \hspace{2mm} \|\mathbf{y}_I+\mathbf{e}_I-H\mathcal{D}_2(\mathbf{y}_I+\mathbf{e}_I)\|_2 \leq \epsilon.
\end{equation}
\end{definition}
\textcolor{blue}{In accordance with common usage in the literature \cite{liu2011false, mo2010false, pasqualetti2013attack, khazraei2022learning, zheng2021resilient}, we adopt two metrics to evaluate the effectiveness and stealthiness of the attack. To quantify the effectiveness, we use the estimation error $\|\mathcal{D}(\mathbf{y}_I)-\mathcal{D}(\mathbf{y}_I+\mathbf{e}_I)\|_2$. To quantify the stealthiness, we use the estimation residual $\|\mathbf{y}_I+\mathbf{e}_I-H\mathcal{D}(\mathbf{y}_I+\mathbf{e}_I)\|_2$, since the estimation process is the primary target of sensor attacks. If the estimation result could be spoofed without triggering an alarm, the corresponding control action is then manipulated maliciously, causing otherwise stable controller to drive the system to an undesired state.}
\begin{remark}\label{rmk:ev_orth}
 Without loss of generality, we assume that $\textsf{supp}(\mathbf{e}_i)\cap\textsf{supp}(\mathbf{v}_i) = \emptyset$. Whenever $\textsf{supp}(\mathbf{e}_i)\cap\textsf{supp}(\mathbf{v}_i) \neq \emptyset$, one could redefine a new attack vector $\mathbf{e}_i+\mathbf{v}_{i_{\textsf{supp}(\mathbf{e}_i)}}$ and new noise vector $\mathbf{v}_{i_{\overline{\textsf{supp}(\mathbf{e}_i)}}}$ that will satisfy this assumption.
\end{remark}
\begin{remark}\label{rmk:success_single_attack}
Given an attack history $\mathbf{e}_{I^{-}}=[\mathbf{e}_{i-T+1}^{\top},\mathbf{e}_{i-T+2}^{\top}, \cdots, \mathbf{e}_{i-1}^{\top}]^{\top}$, the FDIA $\mathbf{e}_i$ at time instance $i$ is said to be $(\alpha, \epsilon)$-successful if $\mathbf{e}_I = [\mathbf{e}_{I^{-}}^{\top},\mathbf{e}_i^{\top}]^{\top}$ satisfies the conditions in \eqref{equ:succ_FDIA}.
\end{remark}

Next, since we target an $\ell_2$ MHE \textcolor{blue}{in \eqref{equ:L2_decoder}}, the conditions in \eqref{equ:succ_FDIA} reduces to
\begin{align}
    \|H^{\dagger}\mathbf{e}_I\|_2  \geq \alpha, \hspace{2mm}
    \|(I-HH^{\dagger})\left(\mathbf{y}_I+\mathbf{e}_I\right)\|_2 \leq \epsilon.
\end{align}
The second inequality can further be simplified as follows:
\begin{align*}
     \|(I-HH^{\dagger})\left(\mathbf{y}_I+\mathbf{e}_I\right)\|_2& =  \|(I-HH^{\dagger})\left(H\mathbf{x}_i+\mathbf{v}_I+\mathbf{e}_I\right)\|_2\\
     &= \|(I-HH^{\dagger})\left(\mathbf{v}_I+\mathbf{e}_I\right)\|_2 \le\epsilon.
\end{align*}
Now, since $\|(I-HH^{\dagger})\mathbf{v}_I\|_2\le\epsilon_v$, using the assumption in Remark~\ref{rmk:ev_orth}, the following conditions are sufficient for the $(\alpha,\epsilon)$ criteria in \eqref{equ:succ_FDIA}:

\begin{equation}\label{equ:succ_suff_con}
    \begin{aligned}
    \|H^{\dagger}\mathbf{e}_I\|_2  \geq \alpha, \hspace{2mm}
    \|(I-HH^{\dagger})\mathbf{e}_I\|_2 \leq \sqrt{\epsilon^2 - \epsilon_v^2} \triangleq \tilde{\epsilon}
    \end{aligned}.
\end{equation}
Let $H$ admit the singular value decomposition
$$ H = \begin{bmatrix}U_1 \hspace{2mm} U_2\end{bmatrix} \begin{bmatrix}\Sigma \\ 0\end{bmatrix} V^{\top},
$$
where $U_1 \in \mathbb{R}^{Tm\times n}, U_2 \in \mathbb{R}^{Tm\times (Tm-n)}$, $\Sigma \in \mathbb{R}^{n\times n}$ is a diagonal matrix composed of all non-zeros singular values and $V \in \mathbb{R}^{n\times n}$. The next result gives a parameterization of successful MH-FDIAs that will be used for subsequent design.
\begin{proposition}
Given $\mathbf{w}_1 \in \mathbb{R}^n, \mathbf{w}_2\in \mathbb{R}^{mT-n}$, the moving-horizon FDIA vector given by 
\begin{equation}\label{equ:successful_FDIA_structure}
    \mathbf{e}_I = U_1\Sigma \mathbf{w}_1+U_2 \mathbf{w}_2
\end{equation}
is $(\|\mathbf{w}_1\|_2, \|\mathbf{w}_2\|_2)$-successful \textcolor{blue}{against the estimator-detector pair $\left(\mathcal{D}(\mathbf{y}_I),\textsf{BDD}(\mathbf{y}_I)\right)$ in \eqref{equ:decoder} and \eqref{equ:BDD}.}
\end{proposition}
\begin{proof}
Since $H$ is full-ranked, $\Sigma$ is invertible. It follows then that $H^{\dagger} = V\Sigma^{-1}U_1^{\top}$ and $I-HH^{\dagger} = I-U_1U_1^{\top}=U_2U_2^{\top}$. Then,
$$\begin{aligned}
 \|H^{\dagger}\mathbf{e}_I\|_2 & = \|V\Sigma^{-1}U_1^{\top}(U_1\Sigma \mathbf{w}_1+U_2 \mathbf{w}_2)\|_2 \\
 & = \|V\mathbf{w}_1\|_2 = \|\mathbf{w}_1\|_2,
\end{aligned}
$$
and
$$ \begin{aligned}
 \|(I-HH^{\dagger})\mathbf{e}_I\|_2 &= \|U_2U_2^{\top}(U_1\Sigma \mathbf{w}_1+U_2 \mathbf{w}_2)\|_2\\
 &= \|U_2\mathbf{w}_2\|_2 = \|\mathbf{w}_2\|_2.
\end{aligned}
$$
Thus $\mathbf{e}_I$ is $(\|\mathbf{w}_1\|_2, \|\mathbf{w}_2\|_2)$-successful.
\end{proof}
\textcolor{blue}{Thus far, we have obtained a structure for a successful FDIA, as defined in \eqref{equ:successful_FDIA_structure}, for a static window of length $T$. However, this structure can only ensure static feasibility for the current time window. Next, we derive conditions for recursive feasibility as the observation window moves. At time $i$, the attackers know their injected historical attacks $\mathbf{e}_{I^-}$, then, according to \eqref{equ:successful_FDIA_structure}, the goal at the current time $i$ is to find $\mathbf{e}_i \in \mathbb{R}^m$ such that} 
\begin{equation}
    \begin{bmatrix}
    \mathbf{e}_{I^-} \\ \mathbf{e}_i
    \end{bmatrix} = U_1\Sigma \mathbf{w}_1+U_2 \mathbf{w}_2
\end{equation}
for some $\mathbf{w}_1 \in \mathbb{R}^n$ with $\|\mathbf{w}_1\|_2 \geq \alpha$ and $\mathbf{w}_2 \in \mathbb{R}^{Tm-n}$ with $\|\mathbf{w}_2\|_2 \leq \tilde{\epsilon}$. It is challenging to specify a priori the values of $\alpha$ and $\tilde{\epsilon}$ that can guarantee the condition above for all time. Fortunately, for the purpose of this paper, we regard the stealthiness of the resulting attack signal more important than its effectiveness. Thus, we employ the strategy of pre-specifying $\tilde{\epsilon}$, then obtaining a time-varying bias $\alpha$ by searching for $\mathbf{w}_1$ whose norm is as big as we can guarantee while satisfying the constraint $\|\mathbf{w}_2\|_2 \leq \tilde{\epsilon}$. Let 
\begin{align}\label{eqn:ws}
\mathbf{w}_1^- = \Sigma^{-1}U_1^{\top}\begin{bmatrix}\mathbf{e}_{I^-} \\\mathbf{0}\end{bmatrix}, \hspace{2mm} \mathbf{w}_2^- = U_2^{\top}\begin{bmatrix}\mathbf{e}_{I^-} \\\mathbf{0}\end{bmatrix},
\end{align}
then $ \begin{bmatrix}\mathbf{e}_{I^-} \\\mathbf{0}\end{bmatrix} = U_1\Sigma \mathbf{w}_1^- + U_2\mathbf{w}_2^-.
$
Consequently, we seek vectors $\mathbf{z}_1 \in \mathbb{R}^n$ and $\mathbf{z}_2 \in \mathbb{R}^{Tm-n}$ that satisfy:
\begin{align}
        \textsf{supp}(U_1\Sigma \mathbf{z}_1+U_2 \mathbf{z}_2) = (T-1)m+\mathcal{T} \triangleq \Bar{\mathcal{T}}, \label{equ:condition_initial_support}\\
        \text{ and }
    \left\|\mathbf{z}_2+\mathbf{w}_2^-\right\|_2 \leq \tilde{\epsilon}, \label{equ:condition_initial_inequality}
\end{align}
where $\Bar{\mathcal{T}}$ is the support of $\begin{bmatrix}\mathbf{0} \\ \mathbf{e}_i\end{bmatrix}$. \textcolor{blue}{Thus far, we obtain a parameterization of the attack at the current time $\begin{bmatrix}\mathbf{0} \\ \mathbf{e}_i\end{bmatrix} \triangleq U_1\Sigma \mathbf{z}_1+U_2 \mathbf{z}_2$. If $\mathbf{z}_1$ and $\mathbf{z}_2$ satisfy the conditions in \eqref{equ:condition_initial_support} and \eqref{equ:condition_initial_inequality}, then, according to Remark~\ref{rmk:success_single_attack}, $\mathbf{e}_i$ is $(\alpha, \tilde{\epsilon})$-successful with the historical attacks $\mathbf{e}_{I^-}$. In other words, the attack $\mathbf{e}_i$ maintains the recursive feasibility when the observation window moves from $I-1$ to $I$.}

Next, let $N = \begin{bmatrix}N_1 \\ N_2\end{bmatrix}$ be a matrix \textcolor{blue}{in the null space of $\begin{bmatrix}U_{1}\Sigma & U_{2}\end{bmatrix}_{\Bar{\mathcal{T}}^c}$} such that
\begin{equation}\label{equ:N_define}
    \begin{bmatrix}U_{1_{\Bar{\mathcal{T}}^c}}\Sigma & U_{2_{\Bar{\mathcal{T}}^c}}\end{bmatrix} N = U_{1_{\Bar{\mathcal{T}}^c}}\Sigma N_1 + U_{2_{\Bar{\mathcal{T}}^c}} N_2 = 0,
\end{equation}
then the inequality constraint in \eqref{equ:condition_initial_inequality} is equivalent to
\begin{equation} \label{equ:condition_middle}
    \|N_2 \mathbf{v} + \mathbf{w}_2^-\|_2\leq \tilde{\epsilon},
\end{equation}
for some vector $\mathbf{v}$ since $\mathbf{z}_2$ must be in the range space of $N_2$. The next result gives a necessary and sufficient condition for \eqref{equ:condition_middle}.

\begin{proposition}
The inequality in \eqref{equ:condition_middle} is feasible if and only if 
\begin{equation}\label{equ:final_condition}
    \|N_2^{\perp^{\top}}\mathbf{w}_2^-\|_2 \leq \tilde{\epsilon}.
\end{equation}
\end{proposition}
\begin{proof}
Using the identity\footnote{Since $N_2$ is defined by equation \eqref{equ:N_define}, one could assume, without loss of generality, that the columns of $N_2$ are orthogonal. Thus, $N_2^{\top} = N_2^{\dagger}$.} $N_2N_2^{\dagger}+N_2^{\perp}N_2^{\perp}{}^{\top} = I$, it follows that:
$$ \begin{aligned}
        \|N_2\mathbf{v}+\mathbf{w}_2^-\|_2^2 &= \|N_2(\mathbf{v}+N_2^{\dagger}\mathbf{w}_2^-)+N_2^{\perp}N_2^{\perp^{\top}}\mathbf{w}_2^-\|_2^2\\
        & = \|N_2(\mathbf{v}+N_2^{\dagger}\mathbf{w}_2^-)\|_2^2 +\|N_2^{\perp}N_2^{\perp^{\top}}\mathbf{w}_2^-\|_2^2 \\
        & = \|N_2(\mathbf{v}+N_2^{\dagger}\mathbf{w}_2^-)\|_2^2 +\|N_2^{\perp^{\top}}\mathbf{w}_2^-\|_2^2.
\end{aligned}
$$
Thus, $\underset{\mathbf{v}}{\min}\|N_2\mathbf{v}+\mathbf{w}_2^-\|_2 = \|N_2^{\perp^{\top}}\mathbf{w}_2^-\|_2$. The result follows by noting that the inequality in \eqref{equ:condition_middle} is feasible if and only if $\displaystyle \min_{\mathbf{v}} \|N_2\mathbf{v}+\mathbf{w}_2^-\|_2 \leq \tilde{\epsilon}$.
\end{proof}
\begin{remark}
The above proposition gives the condition for the existence of a feasible attack vector $\mathbf{e}_i \in \mathbb{R}^m$, at time $i$, as
\begin{equation}\label{equ:condition_for_attack_history}
    \left\|N_{2}^{\perp^{\top}} U_{2}^{T} \begin{bmatrix}\mathbf{e}_{I^-}\\\mathbf{0}\end{bmatrix}\right\|_{2} \leqslant \tilde{\epsilon} .
\end{equation}
If, at the current instant, this condition is violated, we simply set $\mathbf{e}_i=\mathbf{0}$. Alternatively, the condition in \eqref{equ:condition_for_attack_history} can also be used with $\mathbf{e}_{(I+1)^-}$ to establish recursive feasibility for the next time instant.
\end{remark}
For each $\mathbf{v}$ satisfying \eqref{equ:condition_middle}, 
\begin{equation}\label{equ:effectiveness_param}
    \alpha(\mathbf{v}) = \|N_1\mathbf{v}+\mathbf{w}_1^-\|_2
\end{equation}
is the corresponding effectiveness (state bias) induced by the attack vector
\begin{equation}\label{equ:MH_FDIA}
    \mathbf{e}_I = \begin{bmatrix}
    \mathbf{e}_{I^-} \\ \mathbf{e}_i
    \end{bmatrix} = U_1\Sigma(\mathbf{w}_1^- + N_1\mathbf{v})+U_2(\mathbf{w}_2^- + N_2\mathbf{v}).
\end{equation}

Next, we develop an iterative scheme to improve the effectiveness $\alpha(\mathbf{v})$ while satisfying the stealthiness condition in \eqref{equ:condition_middle}. Given a vector $\mathbf{v}_k$ satisfying \eqref{equ:condition_middle} and a positive constant $\lambda$, let
\begin{equation}\label{equ:d_k}
    \mathbf{d}_k = \left\{
     \begin{array}{lr}
     \frac{N_1^{\top}(N_1\mathbf{v}_k+\mathbf{w}_1^-)}{\|N_1^{\top}(N_1\mathbf{v}_k+\mathbf{w}_1^-)\|_2}, \quad\text{if}\hspace{0.2cm} \|N_1^{\top}(N_1\mathbf{v}_k+\mathbf{w}_1^-)\|_2 \geq \tau, \\\\
     N_1^{\top}(N_1\mathbf{v}_k+\mathbf{w}_1^-), \quad\text{otherwise},
     \end{array}\right.   \hspace{2mm} 
\end{equation}
\begin{equation}
         \mathbf{r}_k = N_2\mathbf{v}_k+\mathbf{w}_2^-, \hspace{2mm}
     \hat{\mathbf{d}}_k = \frac{N_2\mathbf{d}_k}{\|N_2\mathbf{d}_k\|_2} ,\hspace{20mm}
\end{equation}
\begin{equation}\label{equ:lambda_k}
     \lambda_k = \left\{
     \begin{array}{lr}
     \lambda, \quad\hspace{19mm}\text{if}\hspace{0.2cm} \|\mathbf{r}_k+\lambda N_2\mathbf{d}_k\|_2 \leq \tilde{\epsilon}, \\\\
    \frac{-\mathbf{r}_k^{\top}\hat{\mathbf{d}}_k+\sqrt{(\mathbf{r}_k^{\top}\hat{\mathbf{d}}_k)^2-\|\mathbf{r}_k\|_2^2+\tilde{\epsilon}^2}}{\|N_2 \mathbf{d}_k\|_2}, \quad\text{otherwise},
     \end{array}\right. \hspace{2mm}
\end{equation}
then the update law on $\mathbf{v}_k$ is given by 
\begin{equation}\label{equ:update_v}
    \mathbf{v}_{k+1} = \mathbf{v}_k+\lambda_k\mathbf{d}_k,
\end{equation}
where $\tau$ is a pre-defined zero-tolerance value. 
 \begin{theorem}\label{Thm:MH_FDIA_main}
 For the update law in \eqref{equ:update_v}, the followings hold:
 \begin{enumerate}
     \item $\mathbf{v}_{k+1}$ satisfies the constraint in \eqref{equ:condition_middle};
     \item $\alpha(\mathbf{v}_{k+1}) \geq \alpha(\mathbf{v}_k)$.
 \end{enumerate}
 \end{theorem}
 \begin{proof}
 \textcolor{blue}{$1)$ According to \eqref{equ:update_v}, it follows}
\begin{equation}\label{equ:first_equ_proof}
    \begin{aligned}
         \|N_2 \mathbf{v}_{k+1} + \mathbf{w}_2^-\|_2^2 &= \|N_2\mathbf{v}_k+\mathbf{w}_2^-+\lambda_k N_2\mathbf{d}_k\|_2^2 \\
 &= \|\mathbf{r}_k+\lambda_k N_2\mathbf{d}_k\|_2^2 \\
 &= \lambda_k^2\|N_2\mathbf{d}_k\|_2^2+2\lambda_k\mathbf{r}_k^{\top}N_2 \mathbf{d}_k+\|\mathbf{r}_k\|_2^2
 \end{aligned}
\end{equation}
 \textcolor{blue}{According to \eqref{equ:lambda_k},} the two choices for $\lambda_k$ imply that either $\lambda_k=\lambda$, which implies that 
 
 $$\|\mathbf{r}_k+\lambda N_2\mathbf{d}_k\|_2\leq \tilde{\epsilon},$$ 
\textcolor{blue}{or $\lambda_k = \frac{-\mathbf{r}_k^{\top}\hat{\mathbf{d}}_k+\sqrt{(\mathbf{r}_k^{\top}\hat{\mathbf{d}}_k)^2-\|\mathbf{r}_k\|_2^2+\tilde{\epsilon}^2}}{\|N_2 \mathbf{d}_k\|_2}$, then substituting it into \eqref{equ:first_equ_proof} yields}
 \begin{equation}\label{equ:square_residual}
     \lambda_k^2\|N_2\mathbf{d}_k\|_2^2+2\lambda_k\mathbf{r}_k^{\top}N_2 \mathbf{d}_k+\|\mathbf{r}_k\|_2^2 = \tilde{\epsilon}^2,
 \end{equation}
 which also implies that $$\|\mathbf{r}_k+\lambda_k N_2\mathbf{d}_k\|_2\leq \tilde{\epsilon}.$$
 
 $2)$ \textcolor{blue}{By following \eqref{equ:effectiveness_param} and \eqref{equ:update_v},}
 $$
 \begin{aligned}
     \alpha(\mathbf{v}_{k+1}) &= \|N_1 \mathbf{v}_{k+1}+\mathbf{w}_1^-\|_2 \\
     &= \|N_1\mathbf{v}_k+\mathbf{w}_1^- +\lambda_k N_1 \mathbf{d}_k\|_2.
 \end{aligned}
 $$
 If $\|N_1^{\top}(N_1\mathbf{v}_k+\mathbf{w}_1^-)\|_2 \geq \tau$, then, \textcolor{blue}{according to \eqref{equ:d_k}, $\mathbf{d}_k =  \frac{N_1^{\top}(N_1\mathbf{v}_k+\mathbf{w}_1^-)}{\|N_1^{\top}(N_1\mathbf{v}_k+\mathbf{w}_1^-)\|_2}$. Thus,} 
 $$ \begin{aligned}
         \alpha(\mathbf{v}_{k+1}) &= \left\|N_1 \mathbf{v}_k+\mathbf{w}_1^- + \lambda_k \frac{N_1 N_1^{\top}(N_1\mathbf{v}_k+\mathbf{w}_1^-)}{ \|N_1^{\top}(N_1\mathbf{v}_k+\mathbf{w}_1^-)\|_2}\right\|_2 \\
         & = \left\|\left(I+\frac{\lambda_k N_1 N_1^{\top}}{\|N_1^{\top}(N_1\mathbf{v}_k+\mathbf{w}_1^-)\|_2}\right)(N_1 \mathbf{v}_k+\mathbf{w}_1^-)\right\|_2 \\
         &\geq \underline{\sigma}\left(I+\frac{\lambda_k N_1 N_1^{\top}}{\|N_1^{\top}(N_1\mathbf{v}_k+\mathbf{w}_1^-)\|_2}\right)  \alpha(\mathbf{v}_k) \geq \alpha(\mathbf{v}_k).
 \end{aligned}
 $$
 If $\|N_1^{\top}(N_1\mathbf{v}_k+\mathbf{w}_1^-)\|_2 < \tau$, then, \textcolor{blue}{according to \eqref{equ:d_k}, $\mathbf{d}_k = N_1^{\top}(N_1\mathbf{v}_k+\mathbf{w}_1^-)$. Thus,
 $$\begin{aligned}
     \alpha(\mathbf{v}_{k+1}) &= \|N_1\mathbf{v}_k+\mathbf{w}_1^- +\lambda_k N_1 N_1^{\top}(N_1\mathbf{v}_k+\mathbf{w}_1^-)\|_2
     \\ 
     & = \|(I+\lambda_k N_1 N_1^{\top}) (N_1\mathbf{v}_k+\mathbf{w}_1^-) \|_2 \\
     &\geq \underline{\sigma}(I+\lambda_k N_1 N_1^{\top})\alpha(\mathbf{v}_k) \\ &\geq \alpha(\mathbf{v}_k).
 \end{aligned}$$
where $\underline{\sigma}(I+\lambda_k N_1 N_1^{\top})$ is the smallest singular value of $I+\lambda_k N_1 N_1^{\top}$ that must be bigger than $1$.}
 \end{proof}

\begin{remark}
For $T=1$, the MH-FDIA problem reduces to the classic FDIA problem,
\begin{equation}\label{equ:static_FDIA}
\begin{aligned}
        \textsf{Maximize} &\hspace{2mm} \|N_1\mathbf{v}\|_2 \\
    \textsf{Subject to}&\hspace{2mm}  \|N_2\mathbf{v}\|_2 \leq \epsilon
\end{aligned}
\end{equation}
for which numerous results exist in the literature using different relaxation techniques \cite{liu2011false} or an approximate direct solution to the non-convex problem using learning-based approaches \cite{khazraei2021learning}.
\end{remark}
\textcolor{blue}{The first conclusion in Theorem~\ref{Thm:MH_FDIA_main} indicates that \eqref{equ:condition_middle}, which is the constraint for guaranteeing recursive feasibility, can always be met during the iterative procedure. The proposed MH-FDIA scheme is summarized in Algorithm~\ref{Algorithm:PGA1}.}
 
\begin{algorithm}
\textbf{Parameters}: $\tau, U_1, U_2, N_1,N_2, \Sigma, \epsilon, M$.

\uppercase\expandafter{\romannumeral1}.\hspace{2mm}Input historical attacks $\mathbf{e}_{I^-}$.

\uppercase\expandafter{\romannumeral2}.\hspace{2mm}
$\mathbf{w}_1^- = \Sigma^{-1}U_1^{\top}\begin{bmatrix}\mathbf{e}_{I^-} \\\mathbf{0}\end{bmatrix}, \hspace{2mm} \mathbf{w}_2^- = U_2^{\top}\begin{bmatrix}\mathbf{e}_{I^-} \\\mathbf{0}\end{bmatrix}.
$

\uppercase\expandafter{\romannumeral3}.\hspace{2mm} Suggest and improve bias update: Set $\mathbf{v}_1 = -N_2^{\dagger}\mathbf{w}_2^-$\\
\textbf{for} $k=1: M+1$

$ \hspace{5mm} \mathbf{v}_{k+1} = \mathbf{v}_k+\lambda_k
$, where $\lambda_k, \mathbf{d}_k$ are given in \eqref{equ:lambda_k} and \eqref{equ:d_k}.

\textbf{end}

\uppercase\expandafter{\romannumeral4}.\hspace{2mm} Output current attack vector
$ \mathbf{e}_i = U_{12}\Sigma(\mathbf{w}_1^-+N_1\mathbf{v}_M) + U_{22}(\mathbf{w}_2^- +N_2\mathbf{v}_M).
$
\caption{Moving-horizon FDIA}
\label{Algorithm:PGA1}
\end{algorithm}
\begin{figure}
    \centering
    \includegraphics[scale=0.7]{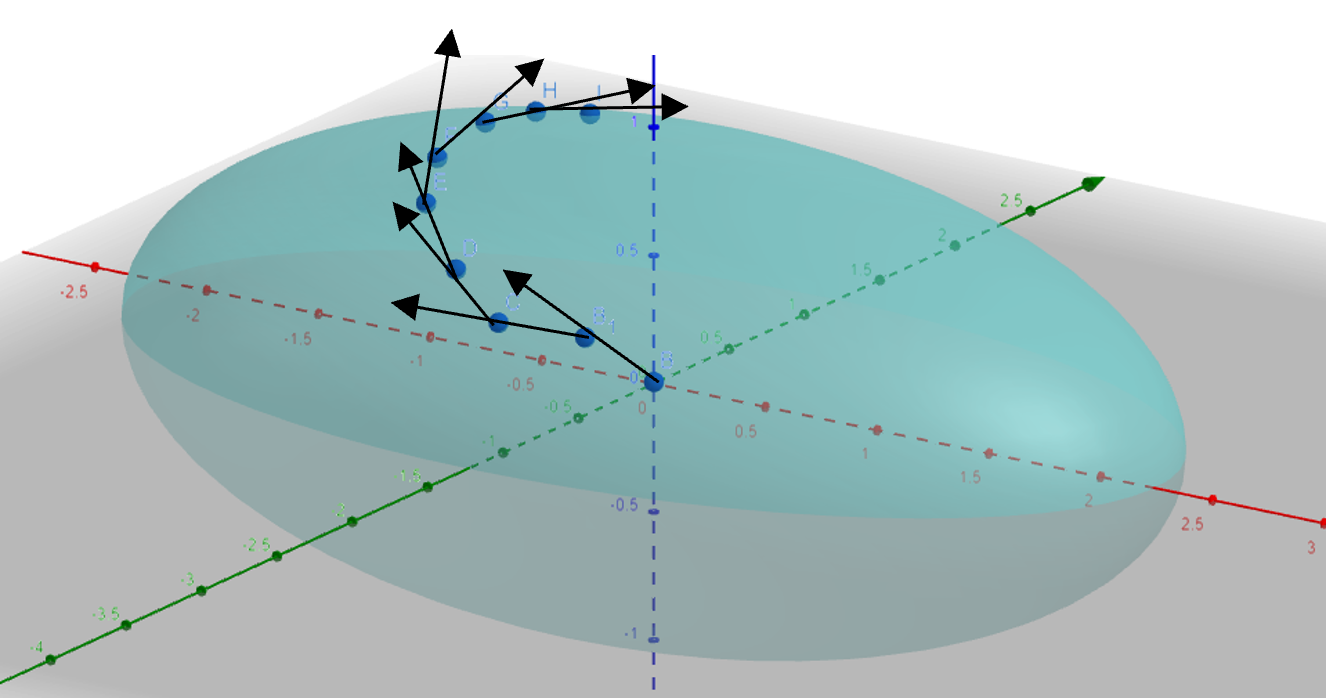}
    \caption{\textcolor{blue}{A conceptual visualization of the Algorithm~\ref{Algorithm:PGA1} for a $3$ dimensional problem}}
    \label{fig:algorithm_visu}
\end{figure}

\textcolor{blue}{In addition, a conceptual visualization of the Algorithm~\ref{Algorithm:PGA1} in $3D$ space is shown in Fig.~\ref{fig:algorithm_visu}. In this space, the constraint in \eqref{equ:condition_middle} could be seen as an ellipsoid. The algorithm searches for a direction that increases the effectiveness $\alpha$ at each iterative step. Furthermore, the update law for step size $\lambda$ in \eqref{equ:lambda_k} ensures that the algorithm eventually terminates on the boundary at time step $M$, as proved in \eqref{equ:square_residual}.}

\section{Simulation \textcolor{blue}{and Experiment}}\label{Sec:Simulation}

\textcolor{blue}{In this section, we validate the efficacy of the MH-FDIA using two case studies: a simulation of a linear regulation control system of the IEEE 14-bus network, and an experiment of a nonlinear path-following control system of a wheeled mobile robot (WMR). Firstly, we validate the theoretical performance of the MH-FDIA and analyze the influence of the parameters of Algorithm 1 on the MH-FDIA's performance using the linear bus system simulation. Through the nonlinear WMR experiment, we showcase the implementation of the proposed MH-FDIA on a nonlinear system and demonstrate its seamless performance in real-world scenarios.
}

\subsection{\textcolor{blue}{Simulation}: IEEE 14-bus system}
\begin{figure}[h!]
    \centering
    \includegraphics[scale=0.4]{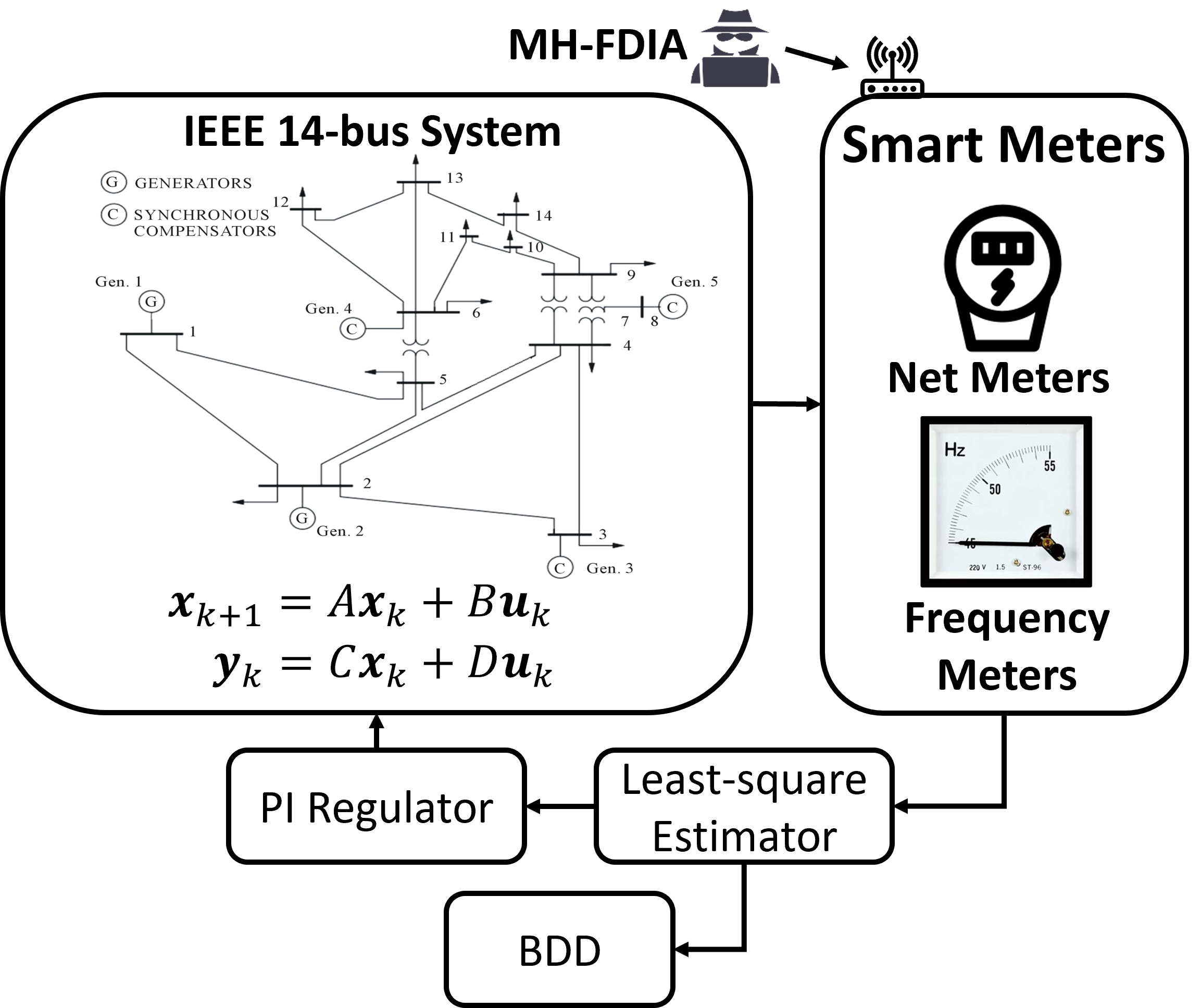}
    \caption{\textcolor{blue}{Schematic description of the simulation of MH-FDIA on the IEEE 14-bus system}}
    \label{fig:sim_power_sys}
\end{figure}

\textcolor{blue}{In this subsection,} the MH-FDIA was implemented on an IEEE 14-bus system with a moving-horizon least-square observer and a residual-based BDD, \textcolor{blue}{as shown in Fig.~\ref{fig:sim_power_sys}.} The MH-FDIA was compared with conventional FDIA designs in the literature.

\textcolor{blue}{A small signal model is constructed by linearizing the generator swing and power flow equations around the operating point \cite{Scholtz2004Observer}. In order to perform linearization, we assume that voltage is tightly controlled at its nominal value; the angular difference between each bus is small; and conductance is negligible therefore the system is lossless.}

\textcolor{blue}{By ordering the buses such that the generator nodes appear first, the admittance-weighted Laplacian matrix can be expressed as $L = \begin{bmatrix}L_{gg} & L_{lg} \\ L_{gl} & L_{ll}\end{bmatrix} \in \mathbb{R}^{N \times N}$, where $N = n_g +n_b$ with $n_g = 5, n_b =14$. Then, by linearizing the dynamical swing equations and algebraic DC power flow equations, we obtain
\begin{equation}\label{equ:bus_system_model}
    \begin{aligned}
   \dot{\mathbf{x}} =& \begin{bmatrix}0 & I \\ -M^{-1}(L_{gg}-L_{gl}L_{ll}^{-1}L_{lg}) & -M^{-1}D_g\end{bmatrix} \mathbf{x} \\
    &+ \hspace{2mm}\begin{bmatrix}0 & 0 \\ M^{-1} & -M^{-1}L_{gl}L_{ll}^{-1}\end{bmatrix} \mathbf{u} ,\\
     \mathbf{y} =& \begin{bmatrix}0 & I \\ - P_{node} L_{ll}^{-1} L{lg} & 0\end{bmatrix}\mathbf{x} + \begin{bmatrix}0 & 0  \\ P_{node} L_{ll}^{-1} & 0\end{bmatrix}\mathbf{u}, \\
    \theta =& -L_{ll}^{-1}(L_{lg}\delta-P_d),
    \end{aligned}
\end{equation} }
where the state variables $\mathbf{x} = [\delta^{\top} \hspace{2mm} \omega^{\top}]^{\top} \in \mathbb{R}^{2 n_g}$ contain the generator rotor angles $\delta  \in \mathbb{R}^{n_g}$ and the generator frequency $\omega \in \mathbb{R}^{n_g}$. The measurements $\mathbf{y} = [\omega^{\top} \hspace{2mm} P_{net}^{\top}]^{\top} \in \mathbb{R}^{n_g+n_b}$ contain the generator frequency $\omega \in \mathbb{R}^{n_g}$ and the net power injected at the buses $P_{net} \in \mathbb{R}^{n_b}$. \textcolor{blue}{$M$ is a diagonal matrix of inertial constants for each generator and $D_g$ is a diagonal matrix of damping coefficients. The stealthiness threshold is set as $5\%$ of the maximal nominal measurement value $\epsilon_i = 0.05  \max(\|\mathbf{y}\|_2) = 0.0318$ for each time step, so the stealthiness threshold for $T$-time horizon is $\epsilon = T \epsilon_i = 0.6352.$} The fixed attack support is $\mathcal{T} = \{1,2,9,11,12,16,17\}$ (about $37$\% of total measurements). The simulation sampling time is set as $T_s = 0.01s$, and the start time for attack injection is $1.8s$. 

 \begin{figure}
\begin{center}
\includegraphics[scale = 0.45]{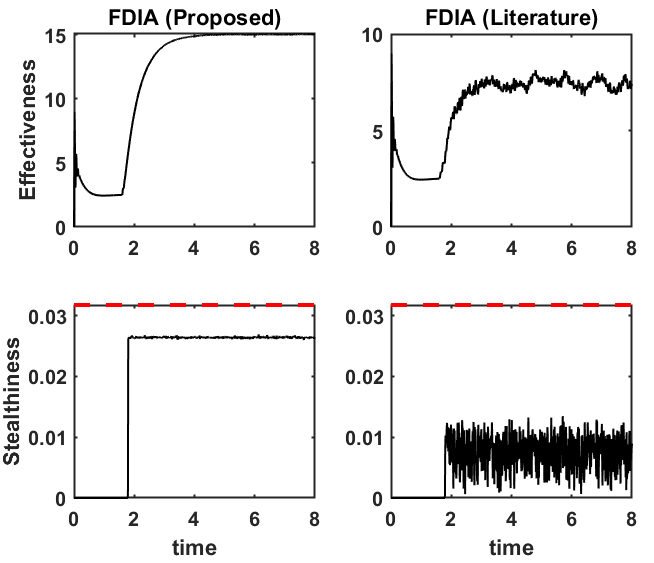}
\end{center}
\caption{A comparison of effectiveness and stealthiness of the static $\ell_2$ FDIA designed in \cite{liu2011false} and the static form of the proposed MH-FDIA against Luenberger observer. (The effectiveness is given by the resulting state estimation error $\|D_2(\mathbf{y}_I)-D_2(\mathbf{y}_I+\mathbf{e}_I)\|_2$. \textcolor{blue}{Red dot line is the threshold of stealthiness, it is $0.0318$ for one time step.)}}
\label{fig:static_compare} 
\end{figure}

\textcolor{blue}{Assuming the attacker has the full knowledge of the system model and parameters in \eqref{equ:bus_system_model},} we compared the MH-FDIA and the eigenvalue-based FDIA \cite{liu2011false} against the Luenberger observer, as well as the moving-horizon least-square observer. Firstly, we implemented the MH-FDIA with a time horizon of $1$, as shown in \eqref{equ:static_FDIA}. Then the MH-FDIA was compared with the static FDIA solved in \cite{liu2011false} against the Luenberger observer. The comparison result is shown in Figure \ref{fig:static_compare}. It is seen that both FDIAs can bypass the detection of the residual-based BDD. However, the proposed MH-FDIA induces a bigger bias in the resulting state estimates, hence is more effective.
\begin{figure}[t!]
\begin{center}
\includegraphics[scale = 0.45]{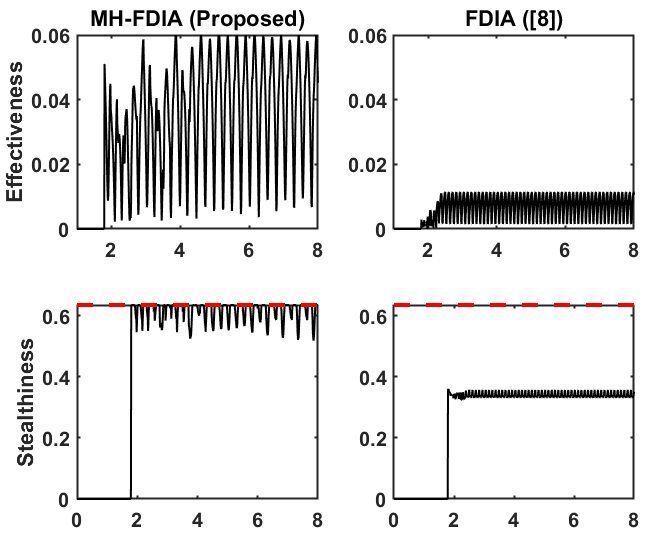}
\end{center}
\caption{A comparison of effectiveness and stealthiness of the proposed MH-FDIA and the one in \eqref{equ:eig_FDIA} against a moving-horizon least-square observer. \textcolor{blue}{(Red dot line is the threshold of stealthiness, it is $0.0318T = 0.6352$ for $T$ time window)}}
\label{fig:MH_compare} 
\end{figure}

Next, we extended the static FDIA in \cite{liu2011false} to a moving-horizon scheme by solving the following eigenvalue problem:
\textcolor{blue}{\begin{equation}\label{equ:eig_FDIA}
        \begin{aligned}
    \textsf{Maximize}_{\lambda, \mathbf{v}} &\hspace{2mm} |\lambda| \\
    \textsf{Subject to}& \hspace{2mm} 
    \left\|\lambda U_{11}\Sigma \mathbf{v} - \mathbf{e}_{I^{-}_{\mathcal{T}_{I^-}}}\right\|_2^2  \leq \tilde{\epsilon}^2
        \end{aligned}
\end{equation}}
where $U_{11} = \begin{bmatrix}I_{n}& 0_{n\times(Tm-n)}\end{bmatrix}U_{I^-}^{\top}$ \textcolor{blue}{corresponds to the direction where $\mathbf{e}_{I^-}$ takes effect on the range space of $H$}. Then the resulting FDIA is given by $\mathbf{e}_{i_{\mathcal{T}_i}} = \lambda^{\star}U_{12}\Sigma \mathbf{v}^{\star}$, where $\lambda^{\star}, \mathbf{v}^{\star}$ are the solution of \eqref{equ:eig_FDIA}, and $U_{12}=\begin{bmatrix}I_{n}& 0_{n\times(Tm-n)}\end{bmatrix}U_i^{\top}$ \textcolor{blue}{corresponds to the direction where $\mathbf{e}_i$ takes effect on the range space of $H$}.

We implemented the proposed MH-FDIA with $\lambda_0 = 1e-4$ and a maximum number of iterations $M = 2000$. The results are shown in Figure \ref{fig:MH_compare}. It is seen that the proposed MH-FDIA has a much bigger effectiveness on the system than the state-of-art counterpart. Moreover, the proposed MH-FDIA explores the whole feasible region, while the one in the literature is more conservative. This indicates that the proposed MH-FDIA algorithm generates more powerful attacks.

\begin{figure}[t!]
\begin{center}
\includegraphics[scale = 0.4]{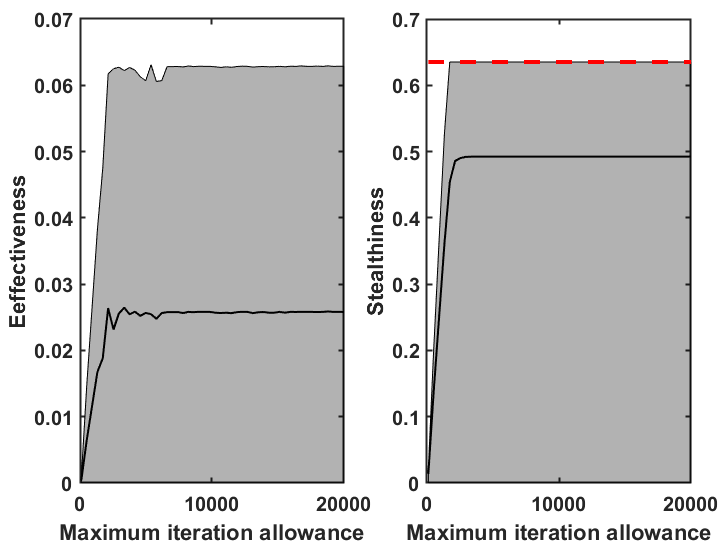}
\end{center}
\caption{The effectiveness and stealthiness of the proposed MH-FDIA with respect to different maximum iteration $M$ in Algorithm~\ref{Algorithm:PGA1}. The black lines are the mean values. The greyed areas show regions of all possible values. \textcolor{blue}{The red dot line is the threshold of stealthiness, it is $0.0318T = 0.6352$ for $T$ time window.}}
\label{fig:perform_vs_iter} 
\end{figure}

Next, we performed additional simulations to explore the effect of the parameters of the proposed algorithm. \textcolor{blue}{We ran the proposed scheme $50$ times with unique values of $M$ ranging from $100$ to $20000$.} The initial step size \textcolor{blue}{was} set as $\lambda = 1e-4$ and the convergence tolerance is set as $1e-4$. $7$ attack channels were chosen at random. \textcolor{blue}{The results in Figure \ref{fig:perform_vs_iter} show the mean effectiveness and stealthiness, along with their corresponding spread for all $50$ simulation cases.} It is seen that the mean effectiveness saturates at $M=2131$ when $\lambda = 1e-4$. The saturation happens sooner ($M=300$) when we increase the step size to $\lambda=1e-3$. This shows that the proposed scheme is computationally efficient and one could adjust the parameters based on available computational resources. 

\begin{figure}[t!]
\begin{center}
\includegraphics[scale = 0.42]{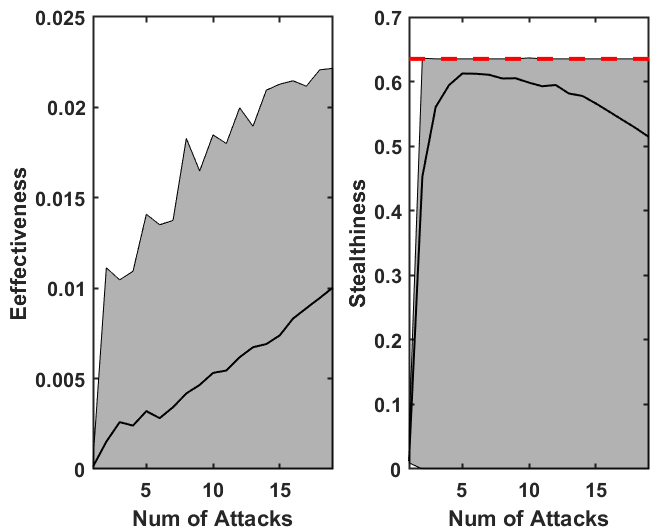}
\end{center}
\caption{The effectiveness and stealthiness of the MH-FDIA for different numbers of attacks. The corresponding attack supports are chosen at random. The black lines are the mean values. The greyed areas show the spread of all possible values. \textcolor{blue}{The red dot line is the threshold of stealthiness, it is $0.0318T = 0.6352$ for $T$ time window.}}
\label{fig:perform_vs_num_attack} 
\end{figure}

Second, we evaluated the performance of the proposed MH-FDIA for different numbers of compromised measurements. $50$ simulation experiments were carried out with randomized attack support for each different number of compromised sensors ranging from $1$ to $19$. We set the initial step size is set as $\lambda = 1e-4$. Figure \ref{fig:perform_vs_num_attack} shows that the effectiveness increases with the number of compromised measurements. On the other hand, all the maximum stealthiness is below the detection threshold. Moreover, it is seen that the maximum and mean stealthiness are close to the detection threshold, which indicates that the proposed algorithm explores the vulnerability space of the system as much as possible. Furthermore, it is seen that the mean stealthiness increases and then decreases since more explorable channels open up, thus making the attacks more stealthy without sacrificing effectiveness.

\subsection{\textcolor{blue}{Experiment:} Wheeled mobile robot}
In this subsection, we implemented the proposed MH-FDIA on a nonlinear path-following control system of a differential-driven wheeled mobile robot (DDWMR). \textcolor{blue}{The attacker is assumed to only have model knowledge of vehicle Kinematics.}
\begin{figure}[ht!]
    \centering
    \includegraphics[scale=0.4]{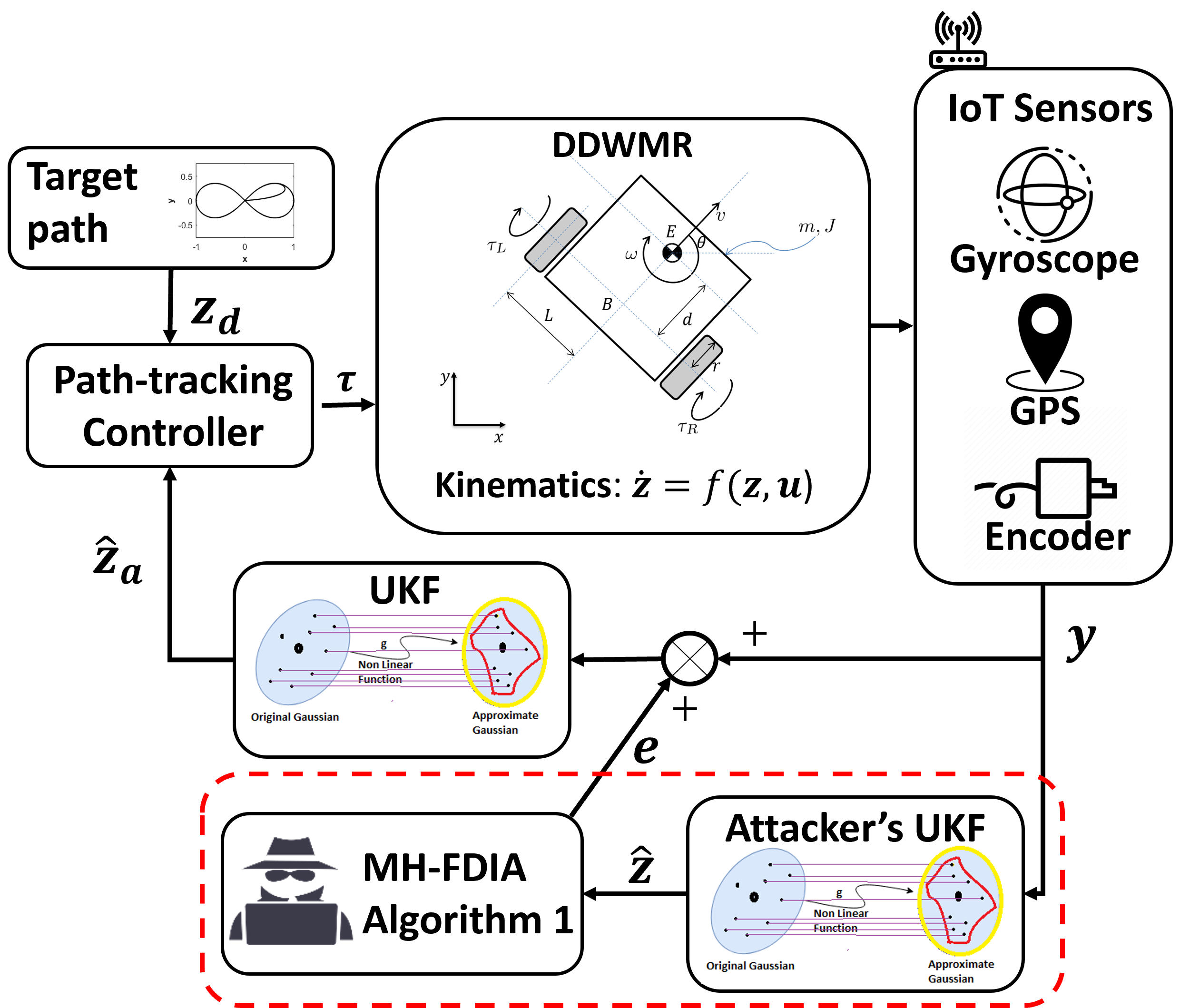}
    \caption{\textcolor{blue}{Schematic description of the simulation of MH-FDIA on the path-tracking control system of DDWMR}}
    \label{fig:DDWMR}
\end{figure}

\textcolor{blue}{The non-holonomic kinematics of DDWMR is given by:
\begin{equation}
    \begin{aligned}
        \begin{bmatrix}
        \dot{\theta}\\
            \dot{\mathbf{z}}
        \end{bmatrix}  &= \begin{bmatrix} 0 & 1 \\ cos(\theta) & -d sin(\theta) \\ sin(\theta) & d cos(\theta) \end{bmatrix} \mathbf{u} + \mathbf{w} \\
        &\triangleq f(\mathbf{x},\mathbf{u}, \mathbf{w}),
    \end{aligned}
\end{equation}
where $\mathbf{u} = [v, \hspace{2mm} \omega]^{\top} \in \mathbb{R}^2$ is the kinematic input vector composing of the longitudinal velocity $v$ $[\textsf{m/s}]$ and the yaw rate $\omega$ $[\textsf{rad/s}]$, $\theta$ $[\textsf{rad}]$ is the heading angle, $\mathbf{z} = [x, \hspace{2mm} y]^{\top} \in \mathbb{R}^2$ is the location vector comprising the $x,y$-coordinates location of the vehicle,  $\mathbf{x} = \begin{bmatrix}\theta & x & y\end{bmatrix}^{\top}$ is the augmented state vector composing of pose state and location states of the vehicle, $\mathbf{w} \sim \mathcal{N}(0,Q)$ is the process noise, and $d = 0.0562m$ is the distance between the medium point of the axis of wheels and center of mass. Given the desired path $\mathbf{z}_d$, we use a common kinematic controller \cite{xie2018trajectory, zheng2020attack}:
\begin{equation}\label{equ:kinematic_WMR}
    \mathbf{u} = \begin{bmatrix}
        cos(\theta) & sin(\theta) \\ -\frac{1}{d}sin(\theta) & \frac{1}{d}cos(\theta)\end{bmatrix} \left(\dot{\mathbf{z}}_d + K(\mathbf{z} - \mathbf{z}_d)\right) ,
\end{equation}
where the control gain $K > 0$.}

The vehicle is equipped with GPS, IMU, and an encoder, resulting in the measurement model
\begin{equation}\label{eqn:meas_model}
    \mathbf{y}  =  \begin{bmatrix} cos(\theta) & 0 \\   0 & sin(\theta) \\  1 & 0 \\  0 & 1 \\  1/4r & L/4r \\  1/4r & -L/4r \end{bmatrix} \cdot \mathbf{z} +\mathbf{v} \triangleq g(\mathbf{x},\mathbf{v})
\end{equation}
where $\mathbf{v} \sim \mathcal{N}(0,R)$. An unscented Kalman filter (UKF) is used to estimate the vehicle's states $\mathbf{x}$. \textcolor{blue}{Firstly, following standard unscented transformation \cite{julier1997new}, we use $2n+1$ sigma points to approximate the state $\mathbf{x}$ with assumed mean $\bar{\mathbf{x}}$ and covariance $P_{\mathbf{x}}$ as follows:
$$
 \begin{aligned}
 &\chi_0= \overline{\mathbf{x}}, \hspace{2mm} \chi_i= \overline{\mathbf{x}} + (\sqrt{(\lambda +n)P_{\mathbf{x}}})_i,\hspace{2mm} i=1,\cdots,n,\\
 &\chi_{i+n}= \overline{\mathbf{x}} + (\sqrt{(\lambda +n)P_{\mathbf{x}}})_{i-n}, \hspace{2mm} i=n+1,\cdots,2n.
 \end{aligned}$$
The corresponding weights for the sigma points are given as $W_0^m = \lambda/(n+\lambda)$, $W_0^c=W_0^m+(1-\alpha^2+\beta)$, $W_i=1/2(n+\lambda)$, and $\lambda=\alpha^2(n+\kappa)-n$ represents how far the sigma points are away from the state, $\kappa \geq 0, \alpha \in (0,1]$, and $\beta=2$ is the optimal choice for Gaussian distribution. Assume $\mathbf{x}_{k-1} \sim \mathcal{N}(\bar{\mathbf{x}}_{k-1}, P_{\mathbf{x},k-1})$, the prediction step is given by
$$
\begin{aligned}
\mathcal{X}_{k} &= f(\mathcal{X}_{k-1},\mathbf{u}_k,\mathbf{w}_k), \hspace{2mm}  \mathcal{Y}_{k} = g(\mathcal{X}_{k}, \mathbf{v}_k),\\
\hat{\mathbf{x}}^{-}_k&=\sum_{i=0}^{2n}W_i \mathcal{X}_{k}, \hspace{2mm} \hat{\mathbf{y}}_{k}^-=\sum_{i=0}^{2n}W_i\mathcal{Y}_{k,i},\\
\hat{P}_{\mathbf{x},k}&=\sum_{i=0}^{2n}W_i(\mathcal{X}_{k}-\hat{\mathbf{x}}_k)(\mathcal{X}_{k}-\hat{\mathbf{x}}_k)^T+R,
\end{aligned}$$
 Next, given the new measurement $\mathbf{y}_k$, the correction step is as follows
\begin{equation}\label{equ:ukf}
    \begin{aligned}
\hat{\mathbf{x}}_{k}&=\hat{\mathbf{x}}_k^-+\mathbf{K}_k(\mathbf{y}_{k}-\hat{\textbf{y}}_{k}^{-}), \hspace{0.2cm} P_{\mathbf{x},k}=\hat{P}_{\mathbf{x},k}-\mathbf{K}_k\hat{P}_{\mathbf{y},k}\mathbf{K}_k^T,
\end{aligned}
\end{equation}
where the Kalman gain is $\mathbf{K}_k=\hat{P}_{\mathbf{z}\mathbf{y}}\hat{P}_{\mathbf{y},k}^{-1}$ with
$$
\begin{aligned}
\hat{P}_{\mathbf{y},k} &=\sum_{i=0}^{2n}W_i(\mathcal{Y}_{k,i}-\hat{\mathbf{y}}_{k}^{-})(\mathcal{Y}_{k,i}-\hat{\mathbf{y}}_{k}^{-})^T+Q, \\
\hat{P}_{\mathbf{z}\mathbf{y}}&=\sum_{i=0}^{2n}W_i(\mathcal{X}_{k,i}-\hat{\mathbf{x}}_k^{-})(\mathcal{Y}_{k,i}-\hat{\mathbf{y}}_{k}^{-})^T.
\end{aligned}
$$ }

\textcolor{blue}{To establish an MH-FDIA generator, we linearized the models of the DDWMR around the equilibrium points $\mathbf{x}_{eq} = [\theta_0 \hspace{2mm} x_0 \hspace{2mm} y_0]^{\top}$, discretized it using Euler's approximation with a sampling time $T_s = 0.01 \textsf{s}$, and iterated forward $T_f=20$ samples:
\begin{equation}\label{equ:linearized_model}
    \mathbf{y}_{I_{T_f}} = H\mathbf{x}_{i-T_f+1} + G\mathbf{u}_{I_{T_f}}
\end{equation}
where 
$$ H = \begin{bmatrix}
            C\\
            C A\\
            C A^2 \\
            \vdots\\
             C A^{T_f}
        \end{bmatrix},
        G = \begin{bmatrix} 0 & 0 & \cdots & 0\\ CB & 0 &\cdots& 0\\ CAB& CB&\cdots & 0 \\ \vdots & \vdots & \ddots & \vdots \\ CA^{T_f-1}B &CA^{T_f-2}B &\cdots & CB \end{bmatrix},
$$
with
\begin{equation*}
\begin{aligned}
A &= \begin{bmatrix}
    1 & 0 & 0\\ 0 & 1 & 0 \\ 0 & 0& 1
\end{bmatrix},\hspace{1mm} B = \begin{bmatrix} 0 & 1 \\ cos(\theta_0) & -d sin(\theta_0) \\ sin(\theta_0) & d cos(\theta_0) \end{bmatrix}T_s, \\
C &= \begin{bmatrix}
        -x_0 sin(\theta_0) & cos(\theta_0) & 0\\
        y_0 cos(\theta_0) & 0 & sin(\theta_0)\\
        0 & 1 & 0\\
        0 & 0 & 1\\
        0 & 1/4r & L/4r \\ 0 & 1/4r & -L/4r\end{bmatrix}.
\end{aligned}
\end{equation*}
Since each state of the kinematic model in \eqref{equ:kinematic_WMR} is an equilibrium point with zero velocity, the attacker must construct their own UKF to obtain the current location states of the vehicle $\mathbf{x}_{eq}$, as shown in Fig~\ref{fig:DDWMR}. Thereafter, the attacker can implement Algorithm \ref{Algorithm:PGA1} with the linearized model in \eqref{equ:linearized_model}.}

\textcolor{blue}{ The MH-FDIA was injected through the third and fourth measurement channels, $\mathcal{T}_i = \{3,4\}$. The stealthiness threshold is set as $\epsilon_i = 0.05$ for each time step, then $\epsilon=T_f \epsilon_i = 1$ for an observation window of length $T_f$. We use the distance from the nominal estimation residual range to measure the stealthiness. The parameters of Algorithm~\ref{Algorithm:PGA1} are chosen as $\lambda_0 = 1e-4$ and $M = 2000$.}

\begin{figure}
    \centering
    \includegraphics[scale=0.5]{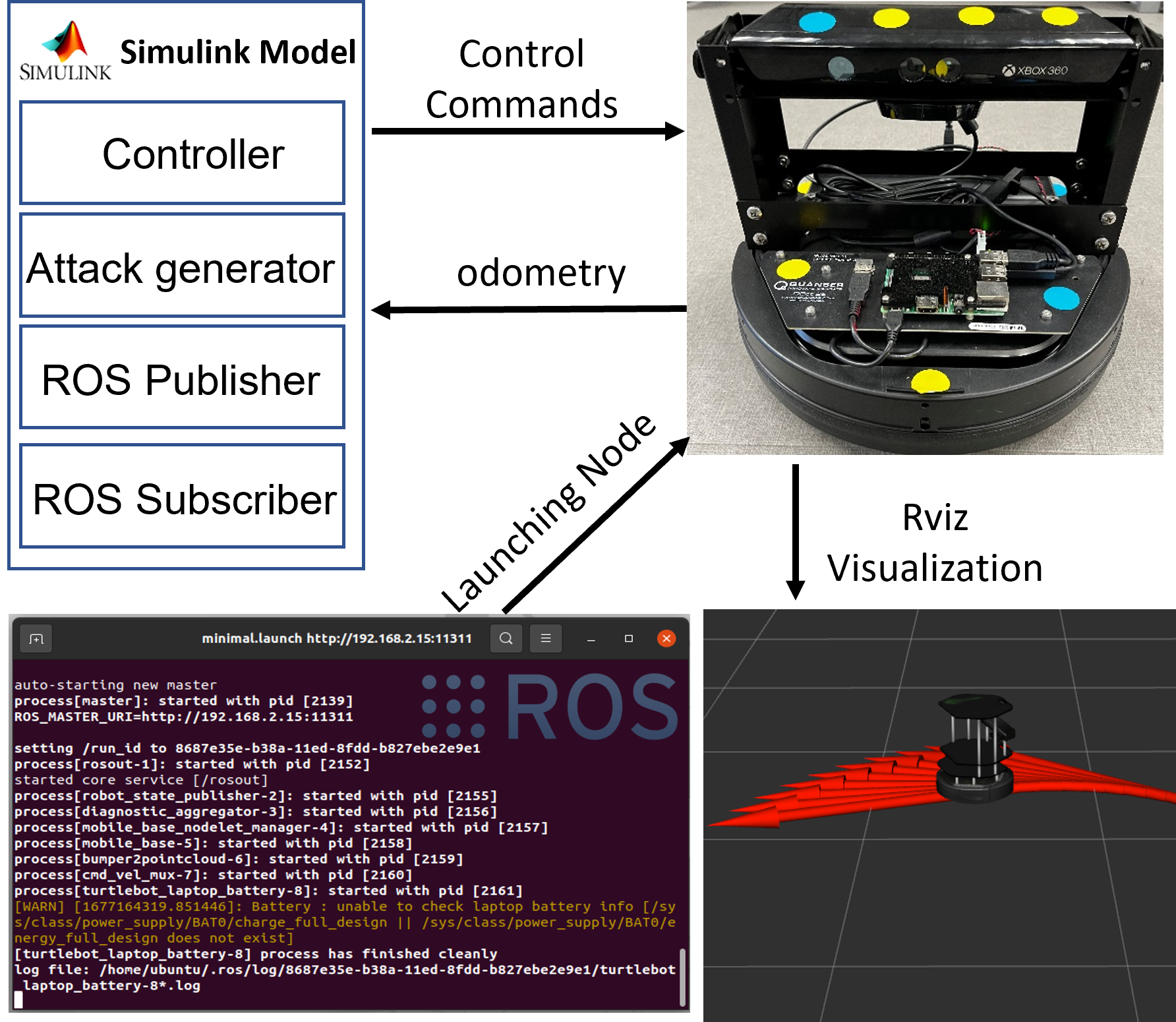}
    \caption{\textcolor{blue}{Experiment platform setup (Simulink + ROS Melodic + Rasberry Pi + DDWMR + Odometry Sensors)}}
    \label{fig:experiment}
\end{figure}

\textcolor{blue}{The experiment platform setup is depicted in Fig.~\ref{fig:experiment}. The DDMWR is constructed using the Yujin 2-wheel Kobuki robot with onboard Rasberry Pi computer and odometry sensors including two-wheel encoders, GPS, and a 3-axis gyroscope. We utilized the Kobuki ROS package\footnote{https://github.com/yujinrobot/kobuki} for the dynamic control of the DDWMR, sensor data collection, and odometry. The path-tracking kinematic controller, UKF and MH-FDIA were all implemented in Matlab/Simulink. The wireless communication channels between Simulink and Rasberry Pi were established based on ROS Melodic.}

\begin{figure}[t!]
    \centering
    \includegraphics[scale=0.4]{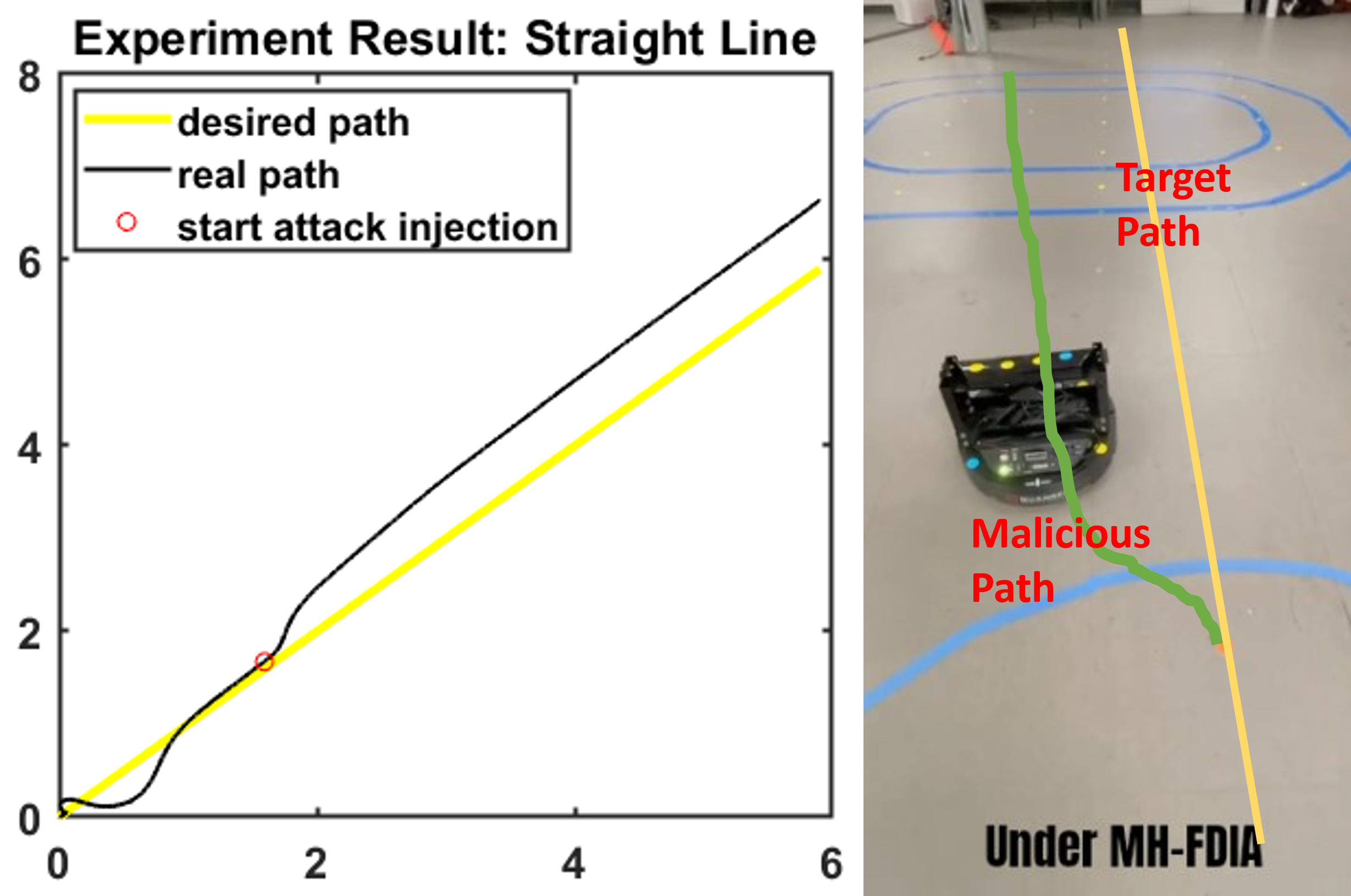}
    \caption{\textcolor{blue}{The path of DDWMR with MH-FDIA in the straight line following test. (Initial pose and location: $\mathbf{x}_0 = \begin{bmatrix}\pi/4 & 0 & 0\end{bmatrix}$)}}
    \label{fig:path_line}
\end{figure}

\begin{figure}[t!]
    \centering
    \includegraphics[scale=0.5]{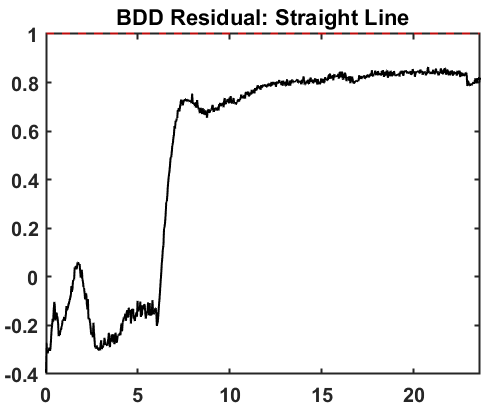}
    \caption{\textcolor{blue}{The BDD residual for the straight line case ($1$ is the threshold, attack injection started at $6$s)}}
    \label{fig:BDD_line}
\end{figure}

\begin{figure}[t!]
    \centering
    \includegraphics[scale=0.45]{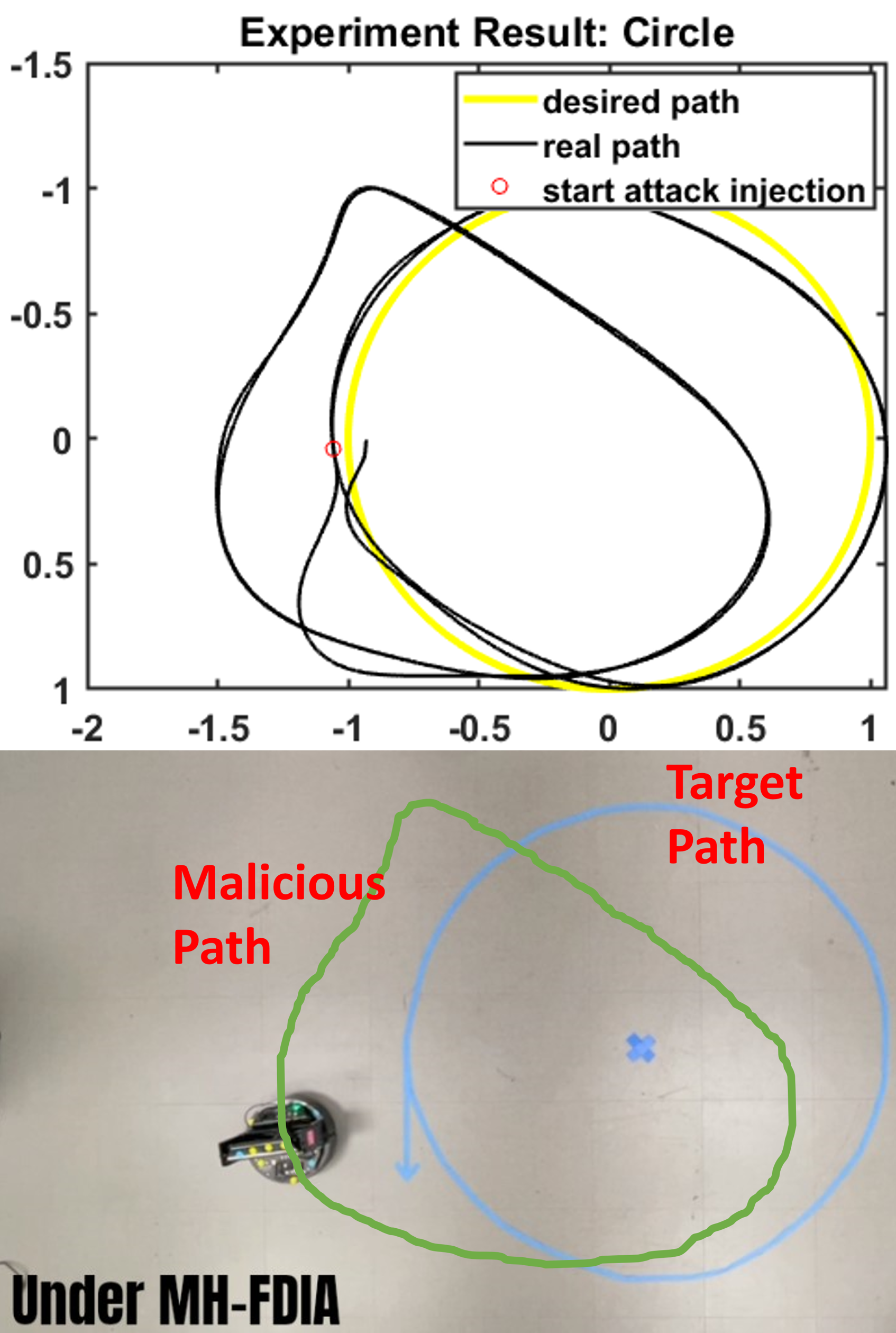}
    \caption{\textcolor{blue}{The path of DDWMR with MH-FDIA in the circular path following test. (Initial pose and location: $\mathbf{x}_0 = \begin{bmatrix}0 & 0 & -1\end{bmatrix}$)}}
    \label{fig:path_circle}
\end{figure}

\begin{figure}[t!]
    \centering
    \includegraphics[scale=0.5]{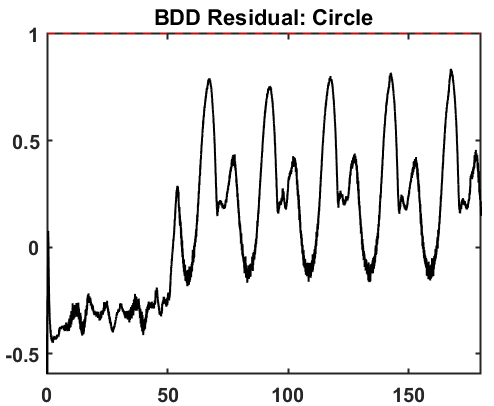}
    \caption{\textcolor{blue}{The BDD residual for the circle case ($1$ is the threshold, attack injection started at $50$s)}}
    \label{fig:BDD_circle}
\end{figure}

\begin{figure}[t!]
    \centering
    \includegraphics[scale=0.45]{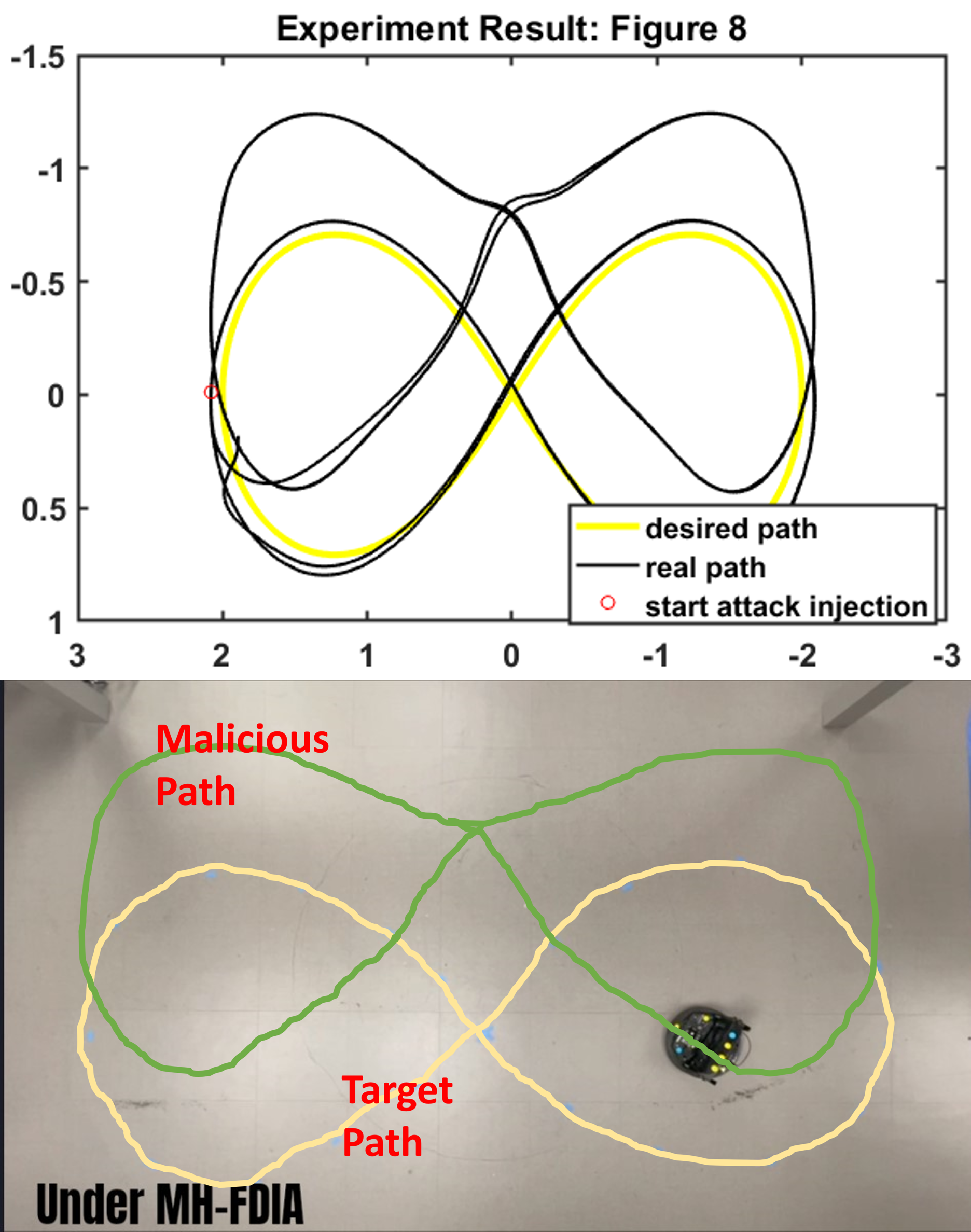}
    \caption{\textcolor{blue}{The path of DDWMR with MH-FDIA in the Figure-8 path following test. (Initial pose and location: $\mathbf{x}_0 = \begin{bmatrix}-\pi/2 & 2 & 0\end{bmatrix}$)}}
    \label{fig:path_eight}
\end{figure}

\begin{figure}[t!]
    \centering
    \includegraphics[scale=0.5]{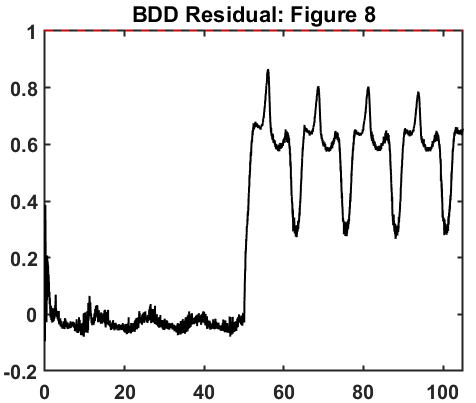}
    \caption{\textcolor{blue}{The BDD residual for the Figure-8 case ($1$ is the threshold, attack injection started at $50$s)}}
    \label{fig:BDD_eight}
\end{figure}

\textcolor{blue}{Three different path-following tasks were undertaken, namely a straight line, a circular path, and a figure-8 path. In the straight line case, the MH-FDIA was initiated at $6$s. Prior to $6s$, the DDWMR was controlled to follow the desired straight line. Following the attack injection, the vehicle gradually deviated from the track, as shown in Fig.~\ref{fig:path_line}. However, the estimation residual was below the threshold $\epsilon$ so the attacks were not detected, as shown in Fig.~\ref{fig:BDD_line}. Similarly, in Fig.~\ref{fig:path_circle} and Fig~\ref{fig:path_eight}, it is seen that the DDWMR was controlled to follow the desired circular path and figure-8 path but run off the track after MH-FDIA was injected. However, the MH-FDIAs were not detected in the two experiments according to Fig~\ref{fig:BDD_circle} and Fig~\ref{fig:BDD_eight}.} It should be noted that the MH-FDIA subtly misled the vehicle to a repeatable wrong path rather than inducing large state estimation errors. \textcolor{blue}{This guarantees that the MH-FDIA is still undetectable in real-world scenarios.} Also, the misled wrong path has a similar shape to the target. This is due to the consideration of historical influence on the current attack injection.

\section{Conclusion}\label{Sec:Conlusion}

In this paper, we demonstrated the importance of historical influence on FDIA design. A systematic MH-FDIA design framework is proposed. Based on a formal definition of successful FDIA, the MH-FDIA design is given against $\ell_2$ MHE and BDD, and shown to be $(\alpha,\epsilon)$-successful. Moreover, an adaptive algorithm is proposed to search for the most successful FDIAs while preserving recursive feasibility. 

 \textcolor{blue}{Developing a MH-FDIA generation algorithm that does not require pre-defined attack support will make the attack generation problem more challenging due to nonconvexity. Developing such algorithm is still an open problem. Additionally, a rigorous quantitative analysis of the MH-FDIA focusing on system's stability is also an interesting question that needs to be addressed. Furthermore, it is also valuable to investigate the applicability of the proposed method in other control systems, such as fuzzy systems and neural network-based systems. Last, to relax the reliance on model knowledge, sample-based techniques could be utilized within the proposed moving-horizon attack generation framework.}


\section{Acknowledgment}
This work is supported in part by the Department of Energy (DOE) under Award Number DE-CR0000005 \textcolor{blue}{and the Defense Advanced Research Projects Agency (DARPA) under Award Number N65236-22-C-8005.}

  \bibliographystyle{elsarticle-num} 
  \bibliography{reference}





\end{document}